\documentclass[final, 3p, 11pt]{elsarticle}
\usepackage{amssymb}
\usepackage{amsmath}

\usepackage[version=4]{mhchem}
\usepackage{makecell}
\usepackage{orcidlink} %for orcid id and link
\usepackage{multirow}
\usepackage{enumitem}
\usepackage{nicefrac}
\usepackage{bm}
\journal{Nuclear Instruments and Methods in Physics Research - section A}

\begin{document}

\providecommand{\tox}{t_{ox}}
\providecommand{\sio}{\ce{SiO2}}
\providecommand{\si}{\ce{Si}}

%% Units
\providecommand{\um}{\mu\text{m}}

%%% table formatting control %%%
% Adds space above text in a cell
%% dont actually use these
\newcommand{\cellpadup}[1]{\rule{0pt}{2ex}#1}

\newcommand{\cellpaddown}[1]{#1\rule[-1ex]{0pt}{2ex}}

\newcommand{\cellpad}[1]{\rule{0pt}{2mm} #1 \rule[-1ex]{0pt}{4ex}}

\newcommand{\degree}{\ensuremath{^\circ}}

% control the 'small' text used in a tablee from here. normalsize > small > footnotesize > scriptsize > tiny
% small table text
\newcommand{\stt}[1]{ {\footnotesize #1} } % make consistent with other tables
\begin{frontmatter}

\title{Demonstrating a broadband Photon Detection Efficiency model on VUV sensitive Silicon Photomultipliers}
% could write mapping, parameterizing, etc

\author[queens]{Austin de St Croix\corref{cor1}\orcidlink{0000-0002-6517-420X}}
\cortext[cor1]{Corresponding Author}
\ead{austindestecroix@gmail.com}

\author[triumf, queens]{Harry Lewis} %% king data-taker, writing
\author[triumf]{Kurtis Raymond} %% modelling, initial work
\author[triumf]{Fabrice Retière} %% bossman
\author[triumf]{Maia Henriksson-Ward} %% MESS, data taking, automotion
\author[princeton]{Giacomo Gallina} %% modeling
\author[triumf, queens]{Nicholas Morrison} %% vera and MESS
\author[triumf]{Aileen Zhang} %% VERA, data-taking

%% other lab people 

\affiliation[queens]{organization={Department of Physics, Engineering Physics and Astronomy, Queen's University},
            addressline={64 Bader Lane},
            city={Kingston},
            postcode={K7L 3N6},
            state={Ontario},
            country={Canada}
}
\affiliation[triumf]{organization={TRIUMF},
            addressline={4004 Wesbrook Mall},
            city={Vancouver},
            postcode={V6T 2A3},
            state={British Columbia},
            country={Canada}
}
\affiliation[sfu]{organization={Department of Physics, Simon Fraser University},
            addressline={8888 University Drive},
            city={Burnaby},
            postcode={V5A 1S6},
            state={British Columbia},
            country={Canada}
}
\affiliation[princeton]{organization={Department of Physics, Princeton University},
            addressline={Jadwin Hall, Washington Road},
            city={Princeton},
            postcode={08544},
            state={New Jersey},
            country={USA}
}

\begin{abstract}
We present a versatile analytic model describing Photon Detection Efficiency (PDE) for P-on-N silicon photomultipliers, with possible applications for device characterization, PDE extrapolation from limited data, simulation and design optimization. Using device specific parameters, SiPM PDE is modeled as a function of wavelength, angle of incidence, voltage, and limited temperature range. By factoring the PDE into transmission and internal efficiency, the performance in liquid nobles and other dense media can be predicted. We present the measurement of the absolute PDE from 350 to 830~nm at 163~K for two VUV sensitive SiPMs: a Hamamatsu VUV4 and Fondazione Bruno Kessler VUV-HD Technology. Additional measurements of relative PDE versus angle are also included. We successfully fit the model to the data, compare with literature and show the model's predictive power by extrapolating PDE to new wavelengths and operation in liquid xenon and argon, which is useful for estimating performance and the impact of external cross-talk in future large-scale experiments. Lastly we use the model to investigate optimizing efficiency for specific applications in astroparticle physics and quantum computing.
\end{abstract}
% \input{absGraphical}

%% Research highlights
% \begin{highlights}
% \item Fit analytic PDE model to broadband data for two VUV sensitive devices
% \item Extrapolated PDE using model, can predict PDE for new devices
% \end{highlights}

%% Keywords
\end{frontmatter}

\section{Introduction}
\label{sec:introduction}
\noindent This work develops a versatile model for the photon detection efficiency (PDE) of silicon photomultipliers (SiPMs), which can be applied across their full range of detectable wavelengths and varied operating conditions. SiPMs have superseded the photomultiplier tube in many physics applications due to benefits including lower operating voltage, radiopurity and compact size. Devices consist of a pixelated array of Single Photon Avalanche Diodes (SPADs), operated in Geiger mode, which produce a characteristic charge pulse in response to a detected photon. SiPMs are used to detect scintillation light from liquid Argon (LAr) and liquid Xenon (LXe) in current or future noble liquid experiments such as SBC, nEXO, Darkside-20K, DUNE, LEGEND and DARWIN/XLZD \cite{sbc_snowmass2021, adhikari_nexo_2021, carnesecchi_darkSide_light_2020, DUNE_updated, legend_result, baudis_darwinxlzd_2024}. A detailed understanding of wavelength and angle dependent PDE is essential for such applications. The model presented here describes SiPM PDE as a function of 5 variables: wavelength, incident angle, overvoltage, operating medium and temperature, with device specific parameters as model inputs. The inputs describe a simplified semiconductor junction structure, avalanche initiation probabilities, and surface optics characteristics, taking optical data from literature as fixed inputs. 

Complete measurement of PDE across this 5-variable parameter space is experimentally challenging; a descriptive model allowing confident extrapolation will improve precision in detector simulations. LAr and LXe emit scintillation light about the VUV wavelengths of 127~nm \cite{fuji_lxe_2015} and 175~nm \cite{Heindl_LAr_2010} respectively, necessitating the use of newly-developed VUV-sensitive SiPMs \cite{baudis_characterisation_2018, gallina_characterization_2019-1, gallina_performance_2022}. In SiPMs the charge avalanche accompanying photon detection produces secondary long-wavelength photons which may escape the device. Measurements show a broad emission spectrum spanning approximately 550~nm into the infrared, with a broad angular distribution\cite{lolx_ext_2025} and an emission yield of order one photon per avalanche \cite{raymond_stimulated_TED_2024}. In a detector these emitted photons may trigger other SiPMs, a process known as external crosstalk (ExCT)\cite{lolx_ext_2025, hampel_optical_2020, joe_emission_2021}. ExCT is expected to be a contributing factor to the overall performance as it deteriorates energy resolution and mimics low occupancy events\cite{Gibbons_2024_flashlight}. As such a key motivation for this work is to accurately predict broadband PDE when submerged in various liquids, using limited angular and spectral vacuum measurements. The optical modeling can be reversed to describe ExCT photons emission, useful for transforming ExCT emission yield from vacuum to dense media \cite{lolx_ext_2025}.

Beyond characterization, the model is intended as a design tool for optimizing SiPMs and single-pixel SPADs, as PDE can be enhanced by tuning optical transmission for particular wavelengths and operating media \autoref{sec:maximize}, as sensitivity to LAr scintillation remains limited in current-generation devices \cite{carnesecchi_darkSide_light_2020}. We explore how efficiencies exceeding 80\% at fixed wavelengths may be achieved for applications in quantum key distribution (QKD), which require efficient single-photon detectors with fast timing resolution \cite{liao_satellite--ground_2017, hadfield_single-photon_2009,ara_shawkat_single_2023}. Lastly, knowledge of avalanche production via electrons or holes informed by this work can inform expectations for dark counts and correlated avalanches in new devices.

The devices studied in this work are two VUV-sensitive SiPMs, the Hamamatsu (HPK) VUV4 and the Fondazione Bruno Kessler (FBK) VUV-HD technology\footnote{This device is referred to as the `VUV-HD3' in other literature}, used in the SBC experiment \cite{sbc_snowmass2021, Herrera_vuv4_2024} and potentially others \cite{Borden_pdeTemp_2024, Wang_2021_cryo_sCurve}. Absolute PDE is measured from 350 to 830~nm at multiple overvoltages, limited by apparatus calibration. After fitting the model, PDE is extrapolated from 120 to 1050~nm and compared with reported VUV efficiencies. Detailed angular dependence of relative PDE is also studied as it pertains to detailed optical modeling, with device optimization explored last.

\begin{table}[h!]
\centering
\caption{Basic characteristics of devices tested here.}\smallskip
\setlength{\tabcolsep}{2.5pt}
\renewcommand{\arraystretch}{1.3}  % Default is 1.0
{\scriptsize
\setlength{\tabcolsep}{5pt}
\renewcommand{\arraystretch}{1.3}
\begin{tabular}{|c|c|c|c|c|}
\hline
\textbf{Device} &
\makecell{\textbf{Fill Factor} \\ (\%)} &
\makecell{\textbf{Pixel Pitch} \\ ($\um$)} &
\makecell{\textbf{Area} \\ (mm$^2$)} &
\makecell{\textbf{Breakdown Voltage} \\ \textbf{at 298~K}(V)} \\
\hline
\makecell{HPK VUV4\\S13371-6050CQ-02}    & 60 & 50 & $4\times (5.95 \times 5.85)$     & 52 \\
FBK VUV-HD  & 80 & 35 & $5.96 \times 5.56$     & 31 \\
\hline
\end{tabular}
}
\label{tab:SiPM_parameters}
\end{table}

%% Need to show some data in here with figures, but still leave some 'true' data for the fitting section

%% possible figures to include %%
% it would be great if we show some of the analysis that gives the resulting PDE
% - finger plot
% - rate plot (unshadowing)
% if you can export the plots to .csv i can format them to match my plots

\section{Experimental setup and methodology}
\label{sec:experimental}
Measurements were taken using the Vacuum Efficiency Reflectivity and Absorption (VERA) setup at TRIUMF, which is described in detail in \cite{Lewis_2025_qy}. A deuterium continuum lamp is used in combination with a vacuum monochromator to provide illumination. A photodiode (Thorlabs FDS10X10) calibrated by NIST from 350 to 830~nm was used to measure the absolute photon flux of the lamp, with absolute efficiency measurements limited to this calibration range. Cooling is provided with liquid nitrogen circulated through a cold finger, with the temperature stability of $\pm 0.1$~K at the nominal measurement temperature of 163~K (LXe temperature). Current measurements were performed using a Keysight B2985A picoammeter. For waveform analysis, SiPM pulses were digitized using a custom-made transimpedance amplifier and CAEN DT5730 digitizer. Breakdown voltage was assessed using the peak of $d(\log{I})/dV$ from IV curve data. Angle of the sample is varied by a stepper motor within a range of motion of $-15^{\circ}$ to $60^{\circ}$ wrt to normal. PDE data is shown throughout \autoref{sec:fitting_the_model} with accompanying fits.

\subsection{Absolute PDE}
\label{sec:measuring_absolutePDE}
The absolute PDE was measured for a range of wavelengths, angles of incidence and overvoltages. Using the adjustable pinhole of the optical collimator, the beam spot ($\leq$ 2~mm) was contained within the active area of the devices, except where otherwise noted. From pulse-counting rates, the light intensity was assessed too weak to cause saturation or non-linearity in device response. For each SiPM the absolute responsivity $R_{SiPM}$ was calculated by measuring the light induced photocurrent on the SiPM and the calibrated diode, as shown in equation \ref{eq:responsivity}. Dark current was subtracted for measurements performed at warmer temperatures.
\begin{equation}
    R_{SiPM}(\lambda) = \frac{I_{SiPM}(\lambda) - I_{D,SiPM}}{I_{Diode}(\lambda) - I_{D,Diode}} \cdot R_{diode}(\lambda)
    \label{eq:responsivity}
\end{equation}
Linear responsivity is then converted to photon detection efficiency using equation \ref{eq:responsivitytoPDE}:
\begin{equation}
    PDE(\lambda) = \frac{R_{SiPM}(\lambda) \cdot E_{photon}(\lambda)}{Q_a} 
    \label{eq:responsivitytoPDE}
\end{equation}
where $E_{photon}$ is the photon energy and $Q_a$ is the gross average charge per single photon detection event. $Q_a$ is the product of charge gain from a single photo-electron (SPE) pulse and the excess charge factor arising from correlated avalanches caused by after-pulsing and internal cross-talk. $Q_a$ is calculated by taking waveform-level data for a given SiPM and determining the rate of `primary' pulses using the method described in \cite{butcher_method_2017}. The SiPM DC current is then measured taken under the same operating conditions, and the current divided by the rate of primary pulses to yield $Q_a$. This measurement is repeated for various overvoltages $V_{ov}$ as $Q_a(V_{ov})$ increases non-linearly with voltage. Although $Q_a$ is not expected to vary with wavelength, measurements were taken in both the dark and under illumination at several wavelengths in order to corroborate the measured values, with uncertainty on $Q_a$ propagated through to the PDE. 

\subsection{Angular dependence of PDE}
\label{sec:measuring_relativePDE}
The relative change in PDE with respect to angle was measured in order to probe the optical properties of device surfaces. The measurement is independent of excess charge factor and therefore was performed using only IV methods. It is also not restricted to the calibration range of the photodiode and was measured from 160 to 850~nm. The DC current at fixed wavelengths was normalized by current at normal incidence, after applying dark current subtraction on warm datasets.

The MC width was varied between some runs. An optical pinhole of either 2~mm or 0.5~mm was used, with a larger pinhole giving higher photocurrent but a more limited range of incidence angles due to beam containment or blocking by device packaging. The maximum achievable angle for the VUV4 devices within VERA is between $30-50\degree$ and slightly higher for the FBK, depending on the pinhole size and exact SiPM positioning.

The spread of wavelengths emitted from the monochromator (MC) is incorporated in the analysis by triangular convolution of the PDE function with respect to wavelength, as expected for the transfer function of a simple diffraction grating based MC. This treatment assumes the input spectrum is flat with respect to wavelength. Particularly at wavelengths near peaks in the lamp spectrum, this assumption is not valid and contributes to systematic uncertainties, which explain differences of results in \autoref{sec_fit:fbk_angular}. The mechanical slit controlling the MC FWHM was varied between 0.5-2~mm, with the wavelength scaling treated as a nuisance parameter in the analysis of the oscillatory FBK $PDE(\theta)$ data. This ellipsometric measurement increased the previous MC scaling value by 31\% to 5.9~nm/mm. The updated wavelength FWHM spans 3-12~nm.

\subsection{Measuring electron avalanche triggering probability}
\label{sec:pe_vs_temp}

The probability that an electron reaching the high-field region of the SiPM will trigger an avalanche, $P_e$, can be measured independent of PDE using the method described in \cite{Lewis_2025_qy}, similar to \cite{Zappala_2016_efficiency}. UV light with sufficiently short absorption length is used to ensure that absorption and subsequent generation of carriers is localized to the p-doped region of the device, and only electron induced avalanches occur. The primary pulse rate under illumination (with dark rate subtracted) is measured at various overvoltages, and normalized by the rate at high overvoltage where $P_e$ is assumed to saturate to unity. This assumption is explicitly tested later in the analysis, driven by the expectation that impact ionization via electrons increases with applied voltage \cite{oldham_avalanche, mcintyre_microplasma, mcintyre_avalanche_1973}. Normalizing by the saturated rate yields $P_e(V)$ as all other PDE related variables are independent of $V_{ov}$. The wavelength condition for complete absorption can be verified by measuring the saturation of pulse rate with overvoltage \cite{gallina_avalanche_2019}, as the probability of hole driven avalanches $P_h(V)$ should not saturate within normal SiPM operating conditions. Preliminary measurements using 380~nm light measured $P_e(V)$ for the HPK device at temperatures from 163-253~K and FBK at 210~K, showing minimal temperature dependence.
 
\section{Modeling the photon detection efficiency}

\begin{figure}[h]
\centering
\includegraphics[width=.35\linewidth]{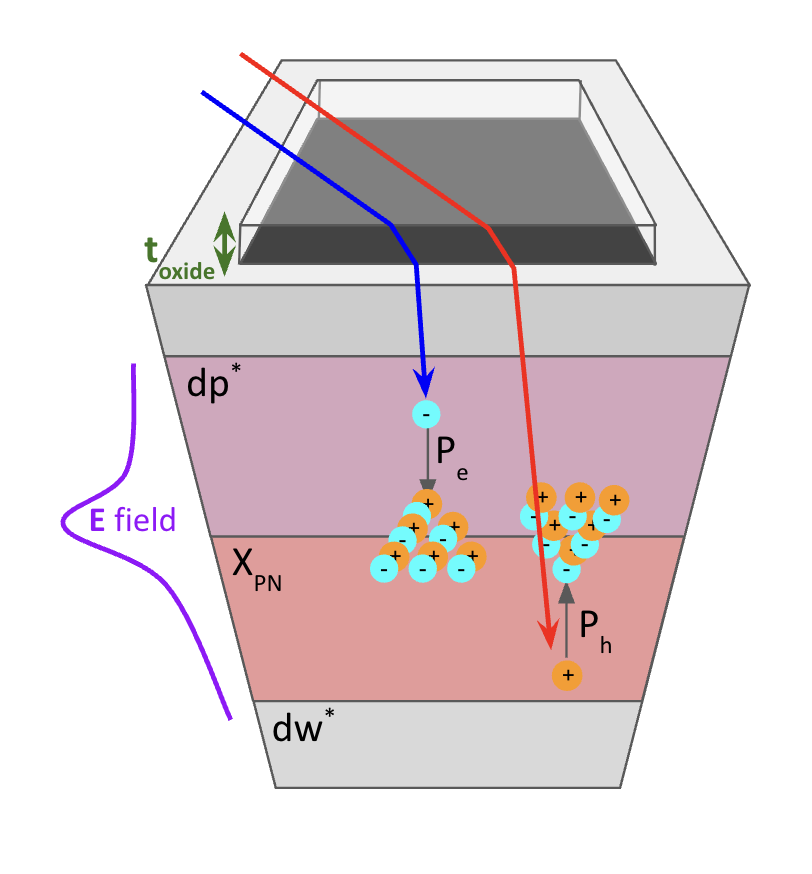}
\includegraphics[width=.6\linewidth]{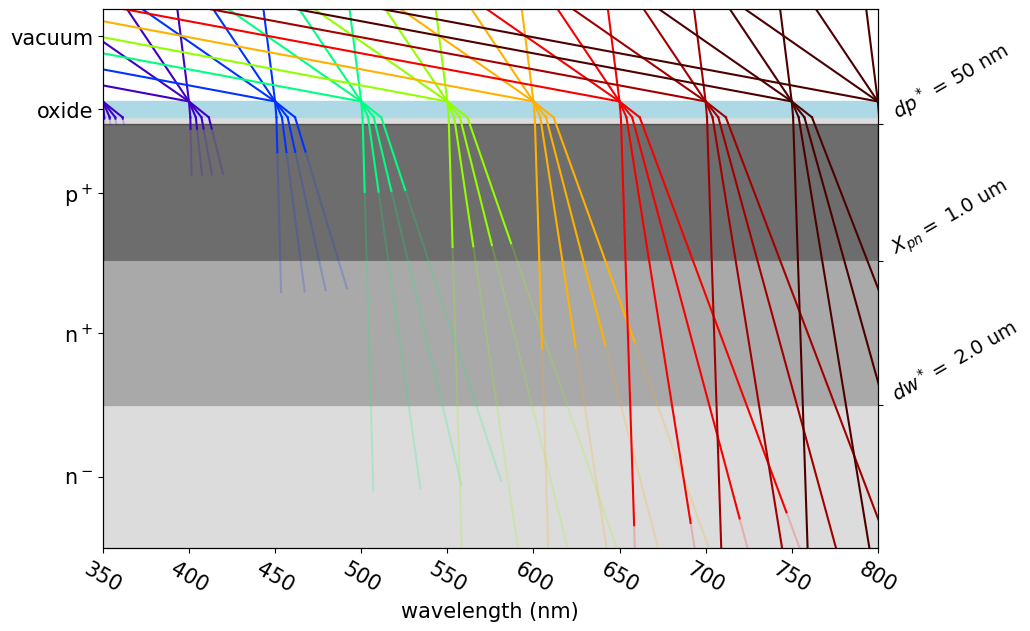}
\caption{\label{fig:pnSchematic}
(Left) Diagram of the modeled SiPM structure, with illustration of the minority carriers drifting into the high field region and initiating avalanches. (Right) Ray diagram of photon paths in SiPM (for real part of internal angle $\theta_3$). Thick lines in silicon correspond to that wavelength's attenuation length $1/\mu$, with fainter lines drawn at length $4/\mu$. $n^-$ denotes the insensitive bulk silicon. The refracted angles are drawn accurately, highlighting the similarity in path length within silicon for different angles of incidence.}
\end{figure}

The experimental variables influencing PDE are photon wavelength ($\lambda$), incidence angle ($\theta$), operating over-voltage ($V$) and to a lesser extent temperature ($T$). Operation in dense media modifies the PDE via optical coupling. We model the PDE using device specific parameters which describe a simplified PN junction\footnote{Naming conventions follow from \cite{gallina_avalanche_2019}. Junction depths are wrt the silicon surface.} and the avalanche triggering probabilities for the corresponding minority carriers. The SiPM surface is modeled as a thin silicon-dioxide layer, assumed sufficient for describing the efficiency of these devices where we infer a p-on-n junction structure. The model can be easily reversed for n-on-p devices. Simplified device structure is shown in \autoref{fig:pnSchematic}. Listed below are model parameters and some defined mathematical quantities:
% ------
% \begin{itemize}
%     \item $\lambda$ - photon wavelength
%     \item $\theta$ - incident photon angle in medium
%     \item $V$ - SiPM operating over-voltage
% \end{itemize}
% ------

% We use many mathematical or derived quantities.
% \begin{itemize}
%     \item $\bar{n} = n+ik$ - the complex refractive index, $k$ being the extinction coefficient
%     \item $\mu = \frac{4\pi k}{\lambda}$ - the attenuation coefficient, dependant on the extinction coefficient $k$
%     \item $\theta_{2, 3}$ - the photon's angle of incidence within the silicon dioxide (2) and bulk silicon (3), derived from
% \end{itemize}

\begin{itemize}[nosep]
    \item $\tox$ - average thickness of the surface silicon dioxide ($\ce{SiO2}$) passivation layer.
    \item $dp^*$ - effective depth representing top of p-doped region; below $dp^*$ electrons are collected and reach the high field region (near $X_{PN}$).
    \item $X_{PN}$ - the center of the p-n junction and location of high field region, boundary where avalanches are either e- or h+ driven.
    \item $dw^*$ - effective depth representing bottom of n-doped region; above $dw^*$ holes are collected and reach the high field region (near $X_{PN}$).
    \item $P_e(V), P_h(V)$ - probability a collected electron, hole triggers an avalanche. Empirically parameterized wrt voltage as $P_e(V;A_e, V_e) = A_e(1-e^{-\frac{V}{V_e}})$ and $P_h(V;A_h, V_h)=A_h(1-e^\frac{-V}{V_h})$.
    \item $FF$ - Fill Factor, fraction of photo-sensitive surface area.
    \item $\mathbf{J}$ - the set of device parameters listed above.
    \item $W_p, W_n$ - probability of photo-absorption within effective p, n regions.
    \item $T_{Si}$ - total probability of transmission into bulk silicon.
    \item $iPDE$ - the internal detection efficiency, efficiency of a photon in bulk silicon.
    \item $H$ - surface structure height $h$ relative to pixel pitch $p$ ($H = h/p$).
\end{itemize}

The PDE equation is factored into the separate terms: $FF$, transmission into silicon $T_{Si}$, and internal efficiency $iPDE$. Internal efficiency is given by the photoabsorption probability in the p (n) region $W_p$ ($W_n$) multiplied by the avalanche triggering probability of the corresponding minority carrier $P_e$ ($P_h$). The collection depths $dp^*$ and $dw^*$ include some diffusive contribution near the edges as described in \cite{gallina_avalanche_2019}; they are effective parameters correlated with the boundaries of the junction's depletion region. The PDE equation is written below with the first line omitting variables and parameters for clarity.
\begin{align}
    PDE & = FF \cdot T \cdot (P_e \cdot W_p + P_h \cdot W_n) \\
    \label{eq:PDE}
    PDE(\lambda, \theta, V; \mathbf{J}) & = FF \cdot T(\theta, \lambda; \tox) \cdot iPDE(V, \lambda; \mathbf{J}) \\
    iPDE(V, \lambda; \mathbf{J}) & = P_e(V) W_{p}(\lambda; dp^*, X_{PN}) + P_h(V) W_{n}(\lambda; X_{PN}, dw^*)
    \label{eq:iPDE}
\end{align}
To an excellent approximation the absorption functions $W_p, W_n$ are independent of incident photon angle $\theta$ due to strong refraction into silicon. The fill factor $FF$ is typically taken as constant, but for precision modeling can vary with surface structure, incident angle or wavelength (see \autoref{sec:ffTheta}, \cite{ACERBI2018_FF}).

\subsection{Input optical data}
\label{sec:inputOpticalData}

\begin{figure}[htbp]
\centering
\includegraphics[width=.9\linewidth]{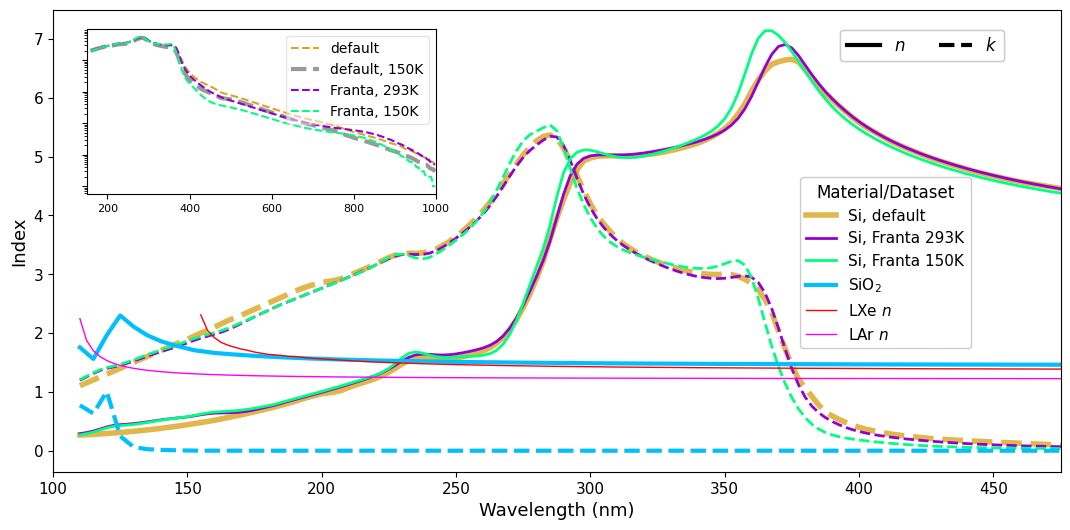}
       
  \caption{\label{fig:input_data} The input optical data used in this work. Solid and dashed lines represent $n$ and $k$, respectively (main figure).  All silicon $n$ values asymptote to $\sim$3.55 at 1000~nm. Inset figure shows Vis-IR $k$ values on a log scale; the differences between these curves drive variation in fit results due to different photoabsorption. The thicker gray `default, 150~K' $k$ curve under the purple line is the default data scaled by the photoabsorption model. Data references are given in the text. The liquid nobles $n$ values are drawn above where $k = 0$.}
\end{figure}

The complex refractive indices $(n, k)$ of silicon, $\sio$ and the detection media are fixed inputs to the model for calculating $T_{Si}$ and $W_p, W_n$. To cover the wavelength range of interest, $\sio$ data was combined from \cite{lithography} over 110-800~nm \footnote{dataset with $k=0$ at $\sim$160~nm selected for consistency with known VUV PDE} and \cite{Malitson} deeper into the IR\footnote{Datasets merged and smoothed in 50~nm region of overlap where datasets had small disagreement}. The $n$ and $k$ data for pure silicon were combined from \cite{lithography} in the VUV (110-200~nm), \cite{nSi_aspnes} (200-750~nm) for UV-Vis\footnote{this dataset is considered more reliable than \cite{nSi_schinke}, which exhibited shape inconsistent with precision Franta dataset \cite{nSi_franta_full_2017} and yielded poorer fit quality} and \cite{nSi_schinke} into the IR. This is referred to as the `default' silicon dataset and used throughout, unless otherwise noted. Curves for ultrapure `float zone' silicon at 150~K are taken from \cite{nSi_franta_full_2017}, which presents a complex theoretical model yielding $n$ and $k$ as a function of silicon purity and temperature\footnote{values taken from \url{https://refractiveindex.info/}}. These values are referred to hereafter as the `Franta' dataset, which are used to investigate model variability with input optical data.

% Some ad-hoc data selection was required; for $\sio$ we chose the dataset from \cite{lithography} with $k=0$ until $\sim$160~nm, consistent with measured VUV PDE and expectation of transparency to LXe scintillation light. For the default $\si$ data, we consider the dataset from \cite{nSi_aspnes} to be more reliable than an alternative dataset from \cite{nSi_schinke}, which exhibited a shape inconsistent with the higher-precision Franta dataset and yielded an overall poorer fit quality.

Input optical data is plotted in \autoref{fig:input_data}, including LXe data from \cite{excitonicBands_lxe_laporte_1977}, \cite{lxe_n_sinnockSmith} and LAr data from \cite{lar_lxe_GRACE2017}. Using pure silicon optical data makes the simplifying assumption that modifications to $n, k$ due to dopants are negligible, avoiding the complexity of including and estimating doping profiles. This assumption is revisited in \autoref{sec:discussion}, as usage of the Franta dataset (with lower $k$ in the NIR) significantly modifies fit parameters. A photoabsorption temperature correction, described in \autoref{model:temp}, does not reduce the room temperature $k$ values of the default dataset as low as the theoretically derived Franta values.

\subsection{Transmittance and reflectivity}
Transmission is modeled as bulk silicon ($\si$) with a thin silicon-dioxide ($\sio$) surface layer, following from the success of this approach in fitting SiPM reflectivity\cite{sipmRef_nexo_guofu}. The vacuum-$\sio$-$\si$ layers are labeled with subscript $1, 2, 3$ respectively. The complex refraction angles in the oxide $\theta_2$ and silicon $\theta_3$ are calculated with Snell's law. \autoref{eq:transmission} is the wave-optics derived transmission from region 1 to 3 for a two interface system with film thickness $\tox$ \cite{kurtis_thesis}, which differs for S($\perp$) or P($\parallel$) polarized photons. $r_{i,j}$, $t_{i,j}$ are the Fresnel coefficients from medium $i$ to $j$. The phase change due to interference within the thin film is given by $\delta$, which drives the oscillatory behavior of $T$ and includes possibly attenuation within the $\sio$. 
\begin{align}
    T_{13, \parallel or \perp} (\lambda, \theta; \tox) &= \frac{\Re (n_3 \cos\theta_3)}{\Re (n_1 \cos\theta_1)} \Big| \frac{t_{12}t_{23}e^{i\delta}}{1 + r_{12}r_{23}e^{i2\delta}} \Big|^2  _{\parallel or \perp}     \label{eq:transmission} \\
    \delta &= 2\pi \frac{n_2 t_{ox}}{\lambda}\cos\theta_2 \\
    T_{Si} &= \nicefrac{1}{2}(T_{13,\parallel} + T_{13, \perp})
\end{align}
The two polarizations are averaged yielding transmission into silicon $T_{Si}$. A similar expression for reflectivity \autoref{eq:Ref_wave} is used for analyzing reflectivity data, where for UV light we assume specular reflectivity from only the active silicon region; $R_{SiPM} \approx FF_o \cdot R_{Si}$. Transmission curves are shown in \autoref{fig:trans_and_abs} and contrasted with the classical ray-optics result. The inset plot shows effects of VUV light absorption in the $\sio$ layer.
\begin{equation}
    R_{13, \parallel or \perp} (\lambda, \theta; \tox)= |\frac{r_{12} + r_{23} e^{i \delta}}{1 + r_{12} r_{23} e^{i \delta}} |^2_{\parallel or \perp}
    \label{eq:Ref_wave}
\end{equation}

% \footnote{the complex valued refractive index is input while only the real component of the resulting angle is kept. The resulting imaginary angle represents the surface of constant amplitude, not constant phase.}

%% Poynting paragraph
An important distinction must be made for P-polarized light at oblique incidence on lossy media such as silicon. In this scenario, the Fresnel field equations predict $T+R < 1.0$. This `missing' Fresnel fraction is typically attributed to absorption within a skin layer, which may contribute to the UV PDE signal. Using the Poynting vector to calculate the transmission retains $R + T = 1.0$ with the direction of energy flow bent further towards the normal than $\Re (\theta_3)$. To account for this, at oblique incidence above 160~nm $T_{Si} = 1-R_{Si}$ is used as \autoref{eq:Ref_wave} is valid over all angular phase space. For oblique angles and $\lambda < 160$~nm, where absorption occurs in the quartz film, disentangling the missing Fresnel fraction from absorption in $\sio$ is less straightforward; using \autoref{eq:transmission} with $\Re ({\theta_3})$ is a good approximation but not exact.
%\autoref{eq:transmission} is valid for normal incidence.

The VUV4 device has a thick quartz window. Fresnel's equations and \sio data is used to calculate the window's transmission and reflectivity $T_W$ and $R_W$, which are combined with $T_{Si}$ and $R_{Si}$ by geometric series giving the total transmission for the system.
\begin{equation}
    \label{eq:T_total}
    T_{total} = \frac{T_W T_{Si}}{1 - R_{Si}R_W}
\end{equation}

\subsection{Internal PDE}
\label{sec:internal_pde}
\begin{figure}[htbp]
\centering
\includegraphics[width=.52\linewidth]{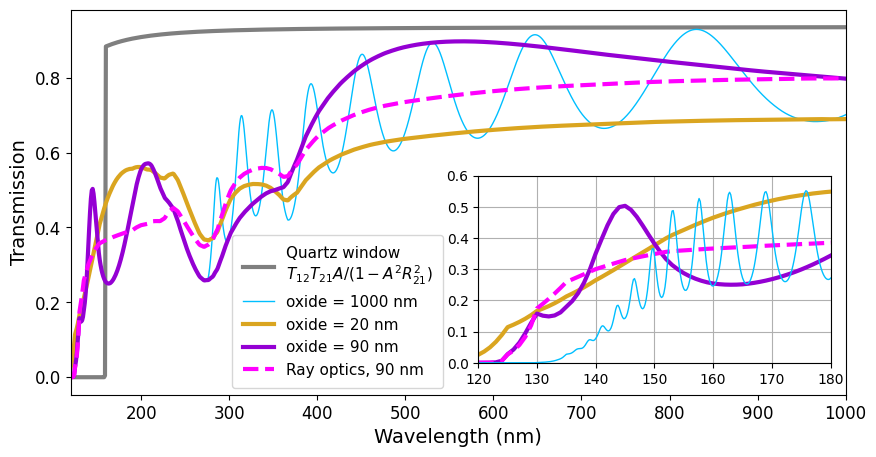}
\includegraphics[width=.47\linewidth]{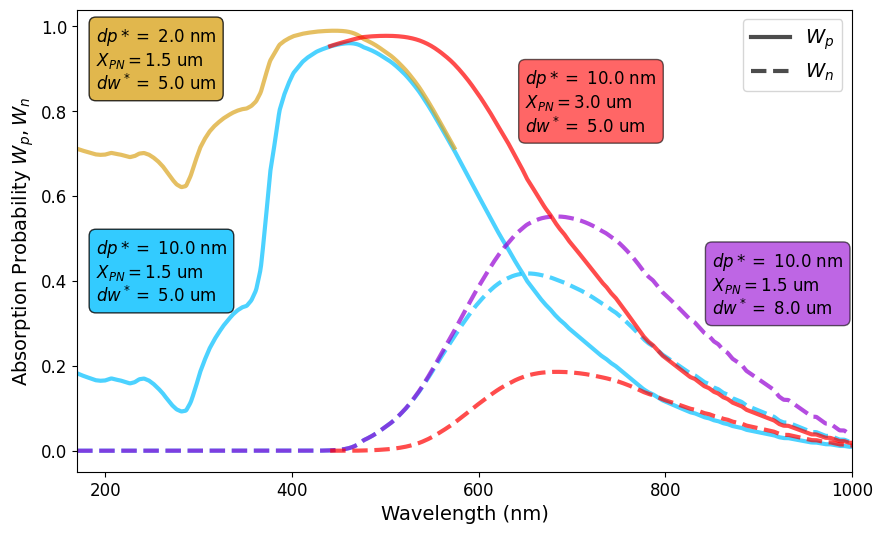}

\caption{\label{fig:trans_and_abs} (Left) Vacuum transmission for the vacuum-quartz-silicon thin film interface in the PDE model. The dashed pink line is the transmission for ray optics; the thin-film transmission (using \autoref{eq:transmission}) oscillates about this mean value. The 20~nm thick (gold) curve is interference-enhanced below 300~nm while the thicker 90~nm curve (purple) oscillates in the VUV and is constructive above 380~nm. The 1$\um$ oxide oscillates rapidly for monochromatic light, and is truncated in VUV for clarity. Inset graph shows the region of interest for liquid noble applications. (Right) Absorption probability vs wavelength for electron ($W_p$, solid) and hole ($W_n$, dashed) collection regions. The different curves highlight the influence of PN geometry.}

\end{figure}

Upon entering silicon a photon can be absorbed in the `dead region'\footnote{so-called because it is assumed photo-generated carriers produced close to the device surface are lost due to recombination} from $0<z<dp^*$, the effective $p$-doped region $dp^*<z<X_{PN}$, effective $n$-doped region $X_{PN}<z<dw^*$ or the insensitive bulk $z > dw^*$. This treatment assumes perfect collection efficiency within these regions and zero outside, and may be referred to as the `hard-edge' model. Silicon absorption length $\mu$ is calculated from the imaginary part of the refractive index.
\begin{equation}
\mu(\lambda; k) = 4 \pi\frac{k}{\lambda}
\label{eq:k_to_mu}
\end{equation}
Photon path lengths are calculated using the real internal angle $\Re(\theta_3)$. Absorption weights for each region, $W_p$ and $W_n$, are the integral over the region scaled by the attenuation prior to (\autoref{eq:w_values}). Curves in \autoref{fig:trans_and_abs} illustrate the effect of different junction geometries; shallower $dp^*$ improves UV absorption, a deeper high-field region $X_{PN}$ shifts the tradeoff between electron or hole driven avalanches, and larger junction ($dw^*$) increases NIR absorption. Due to the strong refraction or `lensing' into the silicon, the $W$ values have negligible dependence on $\theta_1$ over the phase space where transmission remains significant; a NIR photon (least refracted) incident at $80^o$ will have an internal angle of $\sim 20^o$ in bulk. As such $\theta$ dependence is omitted from the internal efficiency $iPDE$.
\begin{align}
    \label{eq:w_values}
    W_p(\lambda, \theta_3; dp^*, X_{PN}) &= e^{-\mu \cdot dp^* / \cos\theta_3}(1-e^{-\mu (X_{PN} - dp^*))/\cos\theta_3}) \\
    W_n(\lambda, \theta_3; X_{PN}, dw^*) &= e^{-\mu \cdot X_{PN} / \cos\theta_3}(1-e^{-\mu (dw^* - X_{PN})/\cos\theta_3})
\end{align}

% \caption{\label{fig:trans_and_abs} Absorption probability vs wavelength for the electron collection region $W_p$ (solid lines) and hole collection region $W_n$ (dashed). The different curves highlight the following influence of PN geometry: proximity to surface ($dp^*$) improves absorption in the UV, a deeper high-field region $X_{PN}$ shifts the tradeoff between electron or hole driven avalanches, and deeper junction ($dw^*$) increases absorption for NIR wavelengths.}

Building upon the study of avalanche triggering probabilities \cite{gallina_avalanche_2019, Zappala_2016_efficiency}, $iPDE$ is given by multiplying the absorption weights by their corresponding minority carrier's avalanche probability, $iPDE = P_e \cdot W_p + P_h \cdot W_n$. This convenient form arises from the assumption that electrons (holes) generated anywhere within the effective p-region (n-region) are successfully drifted to the high-field region about $X_{PN}$ where avalanches are initiated\cite{oldham_avalanche, mcintyre_new_1999} \footnote{defined as the region where impact ionization of both $e^-$ and $h^+$ is above threshold}. The sharp transition from $P_e$ to $P_h$ about $X_{PN}$ stems from the simplification that the high-field region is small relative to the photon absorption scale\footnote{Physically there is a small volume where photo-generated e/h pairs have an avalanche probability somewhere between $P_e$ and $P_h$, as \textit{both} charge carriers pass through a fraction of their respective avalanche inducing regions \cite{chaves_2021_analyticAvalanche}.}. Avalanche efficiency is described with the empirical parameterization $P_e(V) = A_e\cdot(1-e^{-V/V_e})$ and similarly for $P_h$, as demonstrated in \cite{Lewis_2025_qy,Zappala_2016_efficiency}. Later we show $A_e$ can be fixed to 1.0 for the devices studied here. As VUV light generates multiple e-h pairs the $P_e$ term becomes a function of photon energy through the quantum yield $\eta(\lambda) > 1$ \cite{Lewis_2025_qy} . The $P_e$ term then follows binomial statistics $P_e(\lambda) \rightarrow 1 - (1-P_e)^\eta(\lambda)$. 

\subsection{Fill factor and angular dependence}
\label{sec:ffTheta}
The fill factor $FF$ represents the fraction of the SiPM surface sensitive to incident photons. Design elements such as the quenching resistors surrounding each SPAD contribute to the insensitive fraction $1-FF$. For most applications, $FF$ may be treated as a fixed value independent of angle of incidence (AOI) or wavelength. However border effects within bulk modify the electric field near SPAD edges, resulting in possible wavelength dependence of $FF$ \cite{ACERBI2018_FF} for different absorption depths. Large resistor heights relative to SPAD size can shadow the active area effectively reducing $FF$ for oblique incidence. To build a shadowing model, we simplify the geometry and treat the insensitive region as a symmetric, raised square loop surrounding the SPAD. With $p$ the known SPAD pitch, the one dimensional resistor width $ff$ can be calculated from the ratio of sensitive area to total area for a SPAD, $FF = \frac{(p-2ff)^2}{p^2}$. 

Assuming the resistor material is not reflective, for incident light in a plane perpendicular to the resistor the shadowed area is calculated. With a barrier of height $h$ and light at incidence angle $\theta$, the shadow length $h\tan\theta$ is multiplied by the sensitive region's side length $p-2ff$. Defining the relative resistor width $W\equiv ff/p$ and height $H\equiv h/p$, the shadowed fill factor can be written as below, with $FF_o$ the fill factor at normal incidence. In \autoref{sec:fitting_the_model} this model is fit to angular PDE data for the Hamamatsu VUV4 and compared to AFM measurements.
\begin{equation}
    FF(\theta) = F_o - \frac{(p-2ff)h}{p^2}\tan\theta = FF_o - (1-2W)H\tan\theta
    \label{eq-FFtheta}
\end{equation}

\subsection{Temperature and Photoabsorption}
\label{model:temp}
Operating temperature impacts PDE via different mechanisms; modified optics, photoabsorption, carrier mobility and carrier freezeout. Non-monotonic variation in PDE with respect to temperature has been measured, with inflection points at $\sim$150~K \cite{biroth_icasipm_2018} and $\sim$ 200~K \cite{sipm_temp_pde_collazuol_2011}. Studies of a thin device in \cite{acerbi_2023} show further variance of inflection point location and monotonicity depending on wavelength and bias voltage. Negligible temperature respons is reported at 450 nm in \cite{cryogenic_2025_rivera}, and between 160 K and room temperature at 175 nm in vacuum \cite{gallina_performance_2022}. Narrowband studies \cite{garrote_2024_cryoPDE, Wang_2021_cryo_sCurve, Borden_pdeTemp_2024, gu_2025_heliumPDE,PKLightfoot_2008} show varied response. Clearly temperature dependence is wavelength and device dependent. We add a simple extension to the PDE model by including the temperature dependence of photo-absorption for bulk silicon, the expected dominant mechanism for Vis-IR light at temperatures above carrier freezeout. Transmission was minimally impacted when comparing results using the warm or cryogenic $n, k$ input datasets. Carrier freezeout occurs at approximately $\sim 100$~K \cite{biroth_2015_cryogenic}, below the range studied here. A recovery in charge collection efficiency at very low temperatures is the suggested mechanism allowing non-zero PDE at 4~K, which counteracts freezeout effects \cite{biroth_2015_cryogenic, cce_cryogenic_1998}.

% For indirect bandgap materials, photoabsorption decreases with temperature due to both the decrease in phonon population and the increase in the bandgap energy.
As silicon's two lowest bandgaps (1.15, 2.25 eV) are indirect, photoabsorption is strongly temperature dependent from the IR until the direct bandgap at $\sim$ 380~nm (3.4~eV). In \cite{siBandgap_dmPaper} the absorption coefficient $\alpha(\lambda, T) \equiv 1/\mu$ was parameterized in silicon over 440-1000~nm from 4-300~K, for applications in low mass DM searches. We replace the direct bandgap description in \cite{siBandgap_dmPaper} (taken from \cite{siBandgap_solarCells}, \cite{siBandgap_cardona_1985}) with that from \cite{siAbsorption_Bucher} for consistency with photoabsorption values calculated from $k$ data.
% When transforming the parameterized $\alpha$ to $k$ using equation \ref{eq:k_to_mu}, the resonance about 360~nm is not present. Due to this lack of structure, and the systematically low attenuation lengths and sparse wavelength measurements in \cite{siBandgap_dmPaper}, 
To retain the wavelength structure in $k$, which is absent from the sparse data of \cite{siBandgap_dmPaper}, we use the parameterization to calculate the relative change to $\mu(\lambda, T)$ derived from the default $k$ dataset\footnote{Numerical smoothing is done about the direct bandgap threshold where a discontinuity arises. The functional form in selected from \cite{siAbsorption_Bucher} also reduces the discontinuity about this point compared to other functions}. The result is the temperature dependence for photoabsorption in silicon, with $\alpha(\lambda, T)$ the modified form from \cite{siBandgap_dmPaper}:
\begin{equation}
    \mu(\lambda, T; k) = \mu_o(\lambda; k) \frac{\alpha(\lambda, T=300\text{K})}{\alpha(\lambda, T)}
    \label{eq:alpha_relative}
\end{equation}
\section{Experimental Results and Model Fitting}
\label{sec:fitting_the_model}

This section describes strategies and results for fitting the PDE model to data for the Hamamatsu (HPK) VUV4 and FBK VUV-HD SiPMs. We detail the most robust results and impact of modified fitting strategies. The temperature dependent photoabsorption correction (\autoref{model:temp}) is applied to the default optical dataset, and all absolute PDE measurements are performed above 350~nm where the quantum yield is unity \cite{Lewis_2025_qy}.

\subsection{Fitting method}
\label{sec:fit_sequential}
Subsets of the experimental data can constrain PDE parameters in a sequential and fairly independent manner. Junction geometries and corresponding avalanche initiation is separable using wavelengths where either $W_p=0$ or $W_n=0$. Oxide thickness and surface properties can be probed using angular efficiency and/or specular reflectivity. Measurements are tabulated in \autoref{tab:fitting_procedure}.

\begin{table}[ht]
\centering
\caption{\label{tab:fitting_procedure}
Summary of fitting procedures which can constrain different parameters/regions of the PDE model. For thin oxides, reflectivity vs angle is less powerful than vs wavelength.
}
{\smallskip
\footnotesize
\renewcommand{\arraystretch}{2.0}  % Default is 1.0
\begin{tabular}{|c|c|c|c|c|c|}
\hline
\textbf{Order} & \textbf{Physics} & \makecell{\textbf{Constrained} \\ \textbf{parameter}} & \makecell{\textbf{Measurement}} & \textbf{Equation (PDE or other)} & \makecell{\textbf{Data}\\\textbf{Source}} \\ 
\hline
1 & \makecell{Optics \\(thin oxide)} & $\tox$ & UV reflectivity; $\lambda$ & $R(\lambda; \tox)$ (\autoref{eq:Ref_wave}) & \cite{sipmRef_nexo_guofu} \\
\hline
1 & \makecell{Optics \\(thick oxide)} & $\tox$ & Relative $PDE(\theta)$ & \makecell{\rule{0pt}{2ex} $PDE(\theta) \approx PDE(0)~ \frac{T(\theta)}{T(0)}$ \rule{0pt}{2ex}} & \scriptsize{this work}\\
\hline
2 & p depth & $d_p^*$ & \makecell{\scriptsize UV $PDE(\lambda)$\\$\lambda \lesssim 420$~nm, high $V$} & $FF \cdot T \cdot \mathbf{W_p}$ & \scriptsize{this work}\\
\hline
3A & \makecell{e- probability\\ \scriptsize{(PDE independent)}} & $P_e(V)$ & \makecell{\scriptsize SPE rate (V)\\ $\lambda \approx 380$~nm} & Rate$ = A(1-e^{-V/V_e)})$ & \makecell{this work\\ and \cite{Lewis_2025_qy}}\\
\hline
3B & \makecell{p depth and/or\\e- probability} & \scriptsize{$P_e(V)$ and/or $d_p^*$} & \makecell{\scriptsize UV $PDE(\lambda, V)$\\ $\lambda \lesssim 420$~nm} & $FF \cdot T \cdot \mathbf{P_e\cdot W_p}$ & \scriptsize{this work}\\
\hline
4 & \makecell{Junction geo,\\ h+ probability} & \makecell{$X_{PN}, dw^*$, \\ $P_h(V)$} & \makecell{\scriptsize $PDE(\lambda, V)$\\Vis-NIR } & \makecell{$FF \cdot T \cdot ( P_e\mathbf{W_p} + \mathbf{P_h\cdot W_n})$} & \scriptsize{this work}\\
\hline
\end{tabular}
}
\end{table}

Optical transmission is first constrained. Previous measurements \cite{sipmRef_nexo_guofu, raymond_stimulated_TED_2024, kurtis_thesis} indicate the FBK device has a thick oxide layer, permitting $\tox$ to be constrained from oscillations in the angular relative PDE (\autoref{sec_fit:fbk_angular}). Conversely, the absence of oscillations combined with sensitivity to 128~nm light \cite{Pershing_2022_argonPDE, garrote_2024_cryoPDE} indicate the Hamamatsu device has a thin oxide layer ($\lesssim$100~nm). For thin oxides VUV reflectivity provides the best constraint on $\tox$. The effective front of the junction $dp*$ and $P_e(V)$ may be constrained using PDE at UV wavelengths where $W_n \approx 0$. This can be further simplified at high overvoltage where $P_e\approx 1.0$ is assumed. Multiple voltages may be included to constrain $dp^*$ and $P_e(V)$ simultaneously. Lastly visible and IR PDE data probe the remaining junction parameters $X_{PN}$ and $dw^*$ with voltage dependence determining $P_h(V)$. Fitting in this piecewise manner is referred to as the \emph{sequential} method. Results from the sequential fits are used as initial guesses for a refining \emph{global} fit where all parameters are floated simultaneously.

Differences between the sequential and global fits are discussed at the end of this section, intended to inform applications of this model to sparser PDE data. Fits are performed with $P_e, P_h$ floating for each voltage (`nested' fit) or parameterized as $P_{e, h} = A_{e, h}(1-e^{-V/V_{e, h})}$. Alternatively the rate-based $P_e(V)$ measurement (see \autoref{sec:pe_vs_temp}) may be used as in input. Other fitting routines are tested including use of the Franta optical dataset (\autoref{sec:inputOpticalData}).

\subsection{Oxide thickness and Angular PDE for FBK VUV-HD}
\label{sec_fit:fbk_angular}

\begin{figure}[h]
\centering
\includegraphics[width=.9\linewidth]{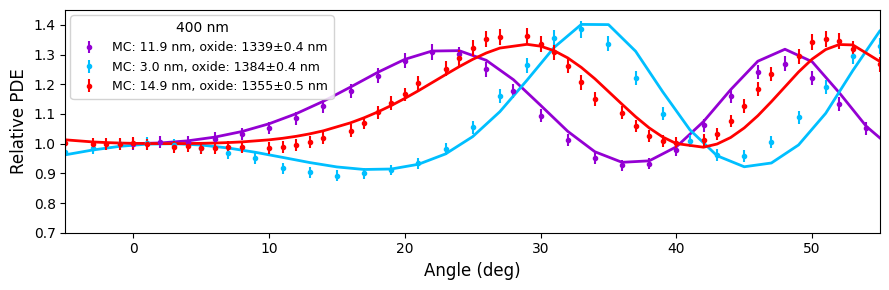}
\caption{\label{fig:fbk_angularData} Three measurements of FBK device's relative PDE versus angle at 400~nm. Monochromator FWHM (MC width) from each campaign and fit oxide thickness $\tox$ are given in the legend. Other wavelengths are omitted for clarity but are included in the fits.}
\end{figure}

The relative angular PDE for the FBK device was measured in three separate campaigns, with monochromator settings modified for each run. Each campaign yielded slightly different oscillation patterns giving different oxide thicknesses within 2\% of the mean 1358~nm. Thicknesses from the Franta input dataset are within 0.5~nm of the default dataset values. Variance between runs is ascribed to the MC transfer function coupled to the underlying lamp spectrum (see \autoref{sec:measuring_relativePDE}), a systematic requiring further investigation for high resolution ellipsometry. \autoref{fig:fbk_angularData} show the measurement at 400~nm although each campaign and fit contain multiple wavelengths.

\subsection{Wavelength-dependent PDE for the FBK VUV-HD}
\begin{figure}[htbp]
\centering
\includegraphics[width=.9\linewidth]{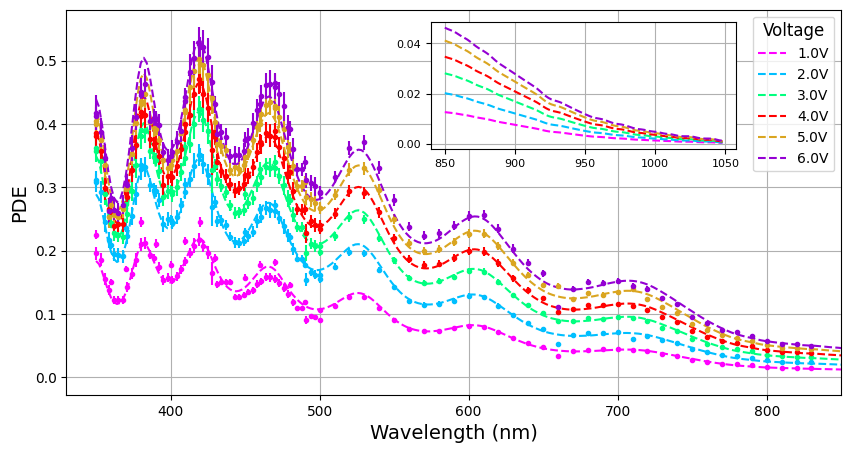}
\caption{\label{fig:fbk_wlenData}
FBK data and the global parameterized fit result using the default input dataset. Outlying data points in 1~V curve from the lower resolution wavelength scan can be seen. Inset plot shows PDE extrapolated to 1050~nm. The resulting $P_e$ curves are almost identical to those for HPK, shown in \autoref{fig:hpk_wlen_fit}.}
\end{figure}

The FBK VUV-HD device's PDE was measured from 350-830~nm at 160~K and 1-6~V overvoltage. Accurate measurement at higher overvoltage was not possible due to an increased rate of correlated avalanches confounding pulse-finding analysis. Data are overlayed with the fit in \autoref{fig:fbk_wlenData}, with inset showing the model extrapolated to longer wavelengths relevant to ExCT. The device parameters derived from the global fits (using the $P_e, P_h$ parameterization) are given in \autoref{tab:fitResults}. The nested $P_e, P_h$ fit produce slightly larger junction geometries. Fitting $PDE(\lambda)$ yielded an oxide thickness of 1358.6~nm, within 1~nm of the angular measurements' mean.

\subsection{Reflectivity and oxide layer thickness for the Hamamatsu VUV4}
\label{sec:HPK_Oxide}
For thin oxides, fitting $\tox$ from PDE is difficult as there is weak angular dependence and wavelength response is convolved with internal efficiency. Stronger constraints are available from VUV specular reflectivity data, reported by the nEXO group for two VUV4 devices with different pixel pitch \cite{sipmRef_nexo_guofu}. We fit the data with \autoref{eq:Ref_wave} multiplied by $FF$. The data in \cite{sipmRef_nexo_guofu} is consistently lower by roughly 20-25\%, but for oxide thickness of approximately $\sim 17$~nm the shape agrees well. Different combinations of wavelength and angular data were fit \footnote{The shadowing effect in \autoref{eq-FFtheta} was included in the angular fit, with an extra factor of two on the $\tan\theta$ term as the specularly reflected light is blocked on entry and exit.}, yielding a $\sim$3~nm thicker oxide on the 75~um pitch device than the 50~um pitch device. Uncertain if this difference is by design or variation in fabrication, we take the weighted average fit values as the central value for $\tox$ with the maximum and minimum as the systematic bounds. The identical procedure is carried out using the Franta optical data, which shifted the central value from 17.2~nm to 19.4~nm. Resulting $\tox$ values are listed in \autoref{tab:fit_optics_oxides}.

\begin{table}[h]
\centering
\caption{HPK oxide thicknesses derived from reflectivity data in \cite{sipmRef_nexo_guofu}, for the given input optical dataset. First two columns are the average per device, last column is weighted average and upper/lower limits.
\label{tab:fit_optics_oxides}}
\smallskip
\footnotesize
\renewcommand{\arraystretch}{1.3}  % Default is 1.0
\begin{tabular}{|c|c|c|c|}
    \hline
    \textbf{Input data} & \textbf{50~$\bm{\um}$ pitch} &  \textbf{75~$\bm{\um}$ pitch} & \textbf{Estimated thickness $\bm{(\tox)}$} \\
    \hline
    default & 16.1~nm & 18.8~nm & \makecell{  $17.2^{+2.4}_{-1.3}$~nm } \\
    % \hline
    % Franta (150~K)  & 17.3~nm & 20.8~nm & \makecell{ $19.4^{+1.9}_{-2.7}$~nm  }\\
    \hline
\end{tabular}
\end{table}

\subsection{Wavelength-dependent PDE for the Hamamatsu VUV4}
\begin{figure}[h]
\centering
\includegraphics[width=0.8\linewidth]{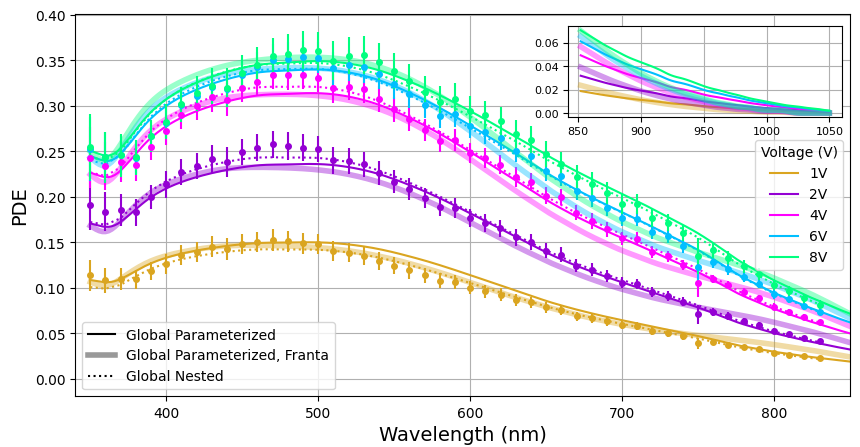}
\caption{
    \label{fig:hpk_wlen_fit} The HPK PDE for selected overvoltages is shown, with 3 fits overlayed. The `nested' (unconstrained) $P_e, P_h$ fits perform slightly better near the $\sim$500~nm maximum. At higher overvoltages the Franta fit (thick line) underestimate the PDE in the NIR. Inset shows both parameterized fits extended to 1050~nm, where the Franta prediction drops faster. Avalanche probabilities are given in \autoref{fig:Ph}.}
\end{figure}

The absolute efficiency of the HPK VUV4 device was measured at 163~K from 350-830~nm for overvoltages 1-8~V. Experimental data with model fit, including various $P_e(V), P_h(V)$ curves, are shown in \autoref{fig:hpk_wlen_fit}. Device parameters derived from the global fit are listed in \autoref{tab:fitResults} while different $P_e, P_h$ results are compared in the discussion. Two other fits are shown; one using the input Franta optical dataset and the unconstrained (nested) fit.

The nested fit yields a lower reduced chi-squared with main improvements at lower overvoltage data. Junction parameters are only slightly modified. The global fit was repeated for the systematic uncertainty on oxide thickness $\tox$, listed as percent difference on the central values in \autoref{tab:fitResults}. All parameters were modified by less than 5\% with the exception of $dp^*$ by -13\%, +43\% for the 15.9~nm and 19.6~nm oxide values, respectively, due to larger shifts in UV transmission.

\subsection{Angular dependence of PDE for the Hamamatsu VUV4 - effects of device microstructure}
\begin{figure}[h]
\centering
\includegraphics[width=0.7\linewidth]{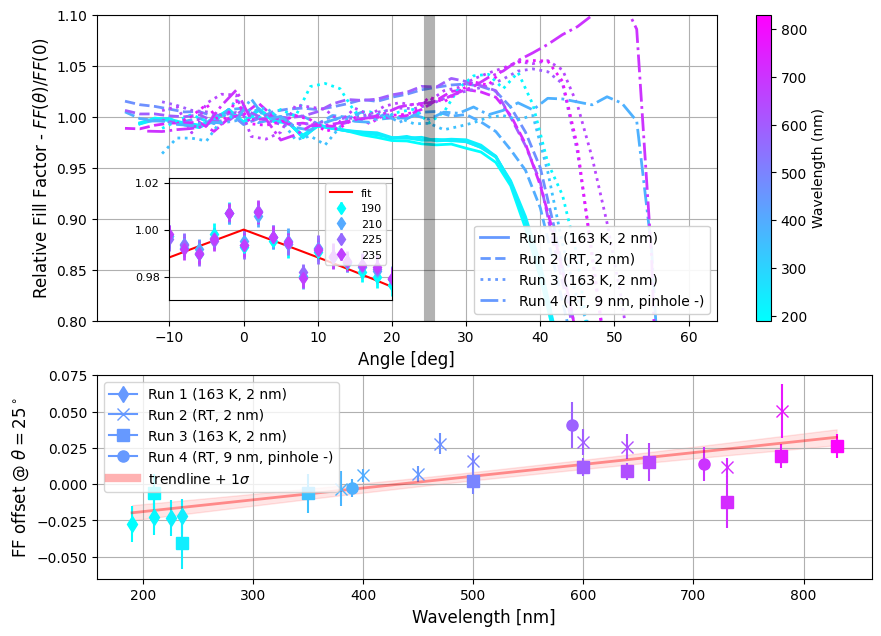}
\caption{Upper plot showing $FF(\theta)/FF(0\degree)$, derived from relative PDE data with transmission factored out. Inset shows the UV data with shadowing \autoref{eq-FFtheta} fit for the effective resistor height. The lower plot shows $FF(\theta)/FF_o - 1$ about 25$\deg$ (dark vertical line) versus wavelength. Some experimental data omitted from upper figure for clarity. \label{fig:hpk_angular_ff}}

    % \caption{Some of the angular data is noisy and some curves are not pictured on the upper plot. The inset shows the shadowing fit of the (non-noisy) UV data. The lower plot shows the offset from unity at 25 degrees, showing linearity with wavelength.
\end{figure}

The angular dependence of PDE was measured for the HPK VUV4 device at wavelengths between 190-830~nm. We observed a small discrepancy between the predicted and measured $PDE(\theta)$, assuming $FF$ is fixed with angle. The relative transmission $T(\theta)/T_o$ is removed from the relative angular PDE (\autoref{sec:measuring_relativePDE}), with the subscript denoting the value at normal incidence and $FF_o = 0.6$. The result is a `measured' relative fill factor, $FF(\theta)/FF_o \approx PDE(\theta)\cdot T_o / PDE_o \cdot T(\theta)$ as the internal PDE is independent of AOI. Relative $FF$ is shown in \autoref{fig:hpk_angular_ff} which is not flat. The falloff at high AOI is due to beam containment or device packaging (onset varies with collimator size). The offset of $FF(\theta)/FF_o$ from unity at $25^\circ$ is plotted versus wavelength, showing a monotonic wavelength dependence; in the UV, the effective $FF(\theta)$ is decreasing with angle while for longer wavelengths $FF(\theta)$ increases.

We assume for UV wavelengths the shadowing model (\autoref{sec:ffTheta}) accurately describes the system, as the polysilicon material is highly absorbing and diffraction should be small\footnote{This was estimated by treating the system as a `half-plane diffraction' problem and integrating the total intensity of E field bent below the plane.}. We fit the UV data (190-235~nm, solid lines in \autoref{fig:hpk_angular_ff}) from -15 to 30 degrees for an effective resistor height $h$ with equation \ref{eq-FFtheta}. The fit yields a value $h = 2.6 \pm 0.1$~um and a reduced chi-squared of 1.6, in excellent agreement with AFM measurements performed by the nEXO collaboration\footnote{Acknowledgements to Prabandha Nakarmi, who performed this measurement in 2022} which report a resistor height between 2.5~um and 3~um. Using the upper and lower values of $\tox$ or the Franta dataset all give resistor heights within this range.

\subsection{Variations with fit procedure}
The assumption that $P_e$ saturates to unity was tested for the HPK device. Floating $A_e$ in the parameterized fits yields $A_e = 0.99\pm 0.01$ and the unparameterized (`nested') fit yields $P_e(8\text{V}) = 0.99 \pm 0.01$. For FBK, $P_e(V)$ does not saturate to 1.0; however the data is limited to 6~V preventing definitive conclusions. The nested fit improves fit quality due to the flexibility in $P_e(V)$. For both devices the sequential and global fit parameters agree, other than $P_e(V)$ when a very narrow UV range is used (350-380~nm).

For the VUV4, variation in $dp^*$ with fit method is driven by the the model's significant curvature about $~\sim$360~nm, a product of transmission and $W_p(\lambda)$. Artificially increasing the monochromator's FWHM could not sufficiently reduce this effect. Including longer wavelength PDE increases $P_e$, at the expense of a worse fit in the UV region where an otherwise shallower $dp^*$ and lower $P_e$ flatten the response. For FBK, the parameterized fit forces a deeper $dp^*$ than the nested fit due to the low overvoltage data; $V_e$ is sensitive to the exact breakdown voltage.

Fitting with the Franta dataset significantly worsens the chi-squared by 2-4 times, and results are omitted from \autoref{tab:fitResults}. The lower photoabsorption and shape reduces the n-type region's size by approximately 2 with a correlated increase in $P_h$ given by $V_h = 4.7\pm 0.3$~V and $V_h = 13.7 \pm 0.4$~V, for HPK and FBK respectively. $dp^*$ is reduced to counteract sharper features in UV optical constants.

\begin{table}[h]
\centering
\caption{A table showing fit results from PDE vs wlen for both devices. FBK oxide thickness $\tox$ is fit from the PDE while the HPK value is independently derived from reflectivity data in \cite{sipmRef_nexo_guofu}. Results listed are from a global fit with $P_e, P_h$ parameterized as $1-e^{-V/V_{e,h}}$, global fit for HPK, and others discussed in the text. The FBK oxide is almost 100 times thicker than HPK, and the junction smaller with a different asymmetric shape. Both devices show similar avalanche probabilities.
\label{tab:fitResults}
}
\smallskip
{\scriptsize
\setlength{\tabcolsep}{2.5pt}
\renewcommand{\arraystretch}{2.5}  % Default is 1.0
\begin{tabular}{|c|c|c|c|c|c|c|c|c|}
    \hline
    \textbf{Device} & \textbf{Fit} & $\bm{\tox}$ (nm) & $\bm{dp^*}$ (nm) & $\bm{X_{PN}}$ (um) & $\bm{dw^*}$ (um) & $\bm{V_e}$ or $\bm{P_e(8\text{V})}$ & $\bm{V_h}$ or $\bm{P_h(8\text{V})}$ & $\bm{\chi^2/\nu}$ \\
    
    \hline

    %% old data w mistake in code
    % \makecell[c]{FBK\\VUV-HD} & Parameterized & 1358.9 $\pm$ 0.2  & $0.90 \pm 0.06$ & $0.51 \pm 0.01$ & $8.39 \pm 0.13$ & $1.83 \pm 0.01$ & $18.90 \pm 0.27$ &  $\nicefrac{1734}{588}$  \\
    \multirow{2}{*}{\makecell[c]{FBK\\VUV-HD}} & Parameterized & 1358.9 $\pm$ 0.2  & $2.73 \pm 0.06$ & $0.42 \pm 0.01$ & $8.03 \pm 0.12$ & $1.87 \pm 0.01$ & $20.1 \pm 0.26$ & $\nicefrac{2703}{588}$  \\
    
    \cline{2-9}
    
     & Nested & 1358.9 $\pm$ 0.2  & $1.1 \pm 0.07$ & $0.50 \pm 0.01$ & $8.43 \pm 0.13$ & $0.88 \pm 0.003$ & $0.23 \pm 0.002$ &  $\nicefrac{1464}{578}$  \\
    \hline
    
    \multirow{2}{*}{\makecell[c]{HPK\\VUV4}} & Parameterized &
    \multirow{2}{*}{\makecell{$17.20^{+2.43}_{-1.31}$}} & 
    \makecell{$1.03 \pm 0.24$ \\ \scriptsize{$+37\%, -19\%$}} &
    \makecell{$2.35 \pm 0.07$ \\ \tiny{$+1.7\%, -0.8\%$}} & 
    \makecell{$13.6 \pm 0.8$ \\ \tiny{$+3.7\%, -1.7\%$}} &
    \makecell{$1.80 \pm 0.022$ \\ \tiny{$+1.7\%, -1.1\%$}} & 
    \makecell{$16.76 \pm 1.21$ \\ \tiny{$+5.1\%, -2.4\%$}} & $\nicefrac{188}{387}$ \\
    \cline{2-2} \cline{4-9}
    
     & Nested &  & 
    $1.31 \pm 0.26$ & $2.08 \pm 0.08$ & 
    $12.4 \pm 0.6$ & $0.99 \pm 0.01$ & $0.41\pm 0.01$ & $\nicefrac{107}{373}$ \\
    
    \hline
    
\end{tabular}
} % end of text size
\end{table}

%% Pe values dont agree perfectly. we learned that you have to know your Vbr `bias' well to use step 3A. Still no Ve values are as low as fits allow so our other PDE terms probably high

% ---- other fit stuff ----
% Allowing $A_e = 0.85$ compensates for overestimate in UV, providing a better fit but is not truly physical; UV $PDE(V)$ matches the shape of the independently measured $P_e(V)$ which rises above 6~V. As such, $P_e$ saturates to a value at least higher than $P_e(6V)=0.88$. Decreasing the $FF$ brings fit results into better agreement with $P_e$ measurements and expectation $A_e \rightarrow 1.0$.

% For either device results from the global and sequential fits agree, with the caveat that the UV wavelength's upper limit is 420~nm (HPK) and 400~nm (FBK) to fit $dp^*$ and $V_e$ (step 3B in \autoref{tab:fitting_procedure}). For HPK restricting the UV fit to 350-380~nm retains agreement other than $dp^*$ coming into tension. Setting $P_e = 1.0$ (step 2 in \autoref{tab:fitting_procedure}) increases $dp^*$ out of agreement.

% Variance in $dp^*$ between procedures means the $P_e$ parameterization is inaccurate for lower voltages (meaning the high-V $dp^*$ is more accurate) or inaccuracy of the predicted UV transmission (all-V $dp^*$ is more accurate).
\section{Model validation and extrapolation for Hamamatsu VUV4}
\label{sec:extrapolation}

\begin{figure}[h]
\centering
\includegraphics[width=.8\linewidth]{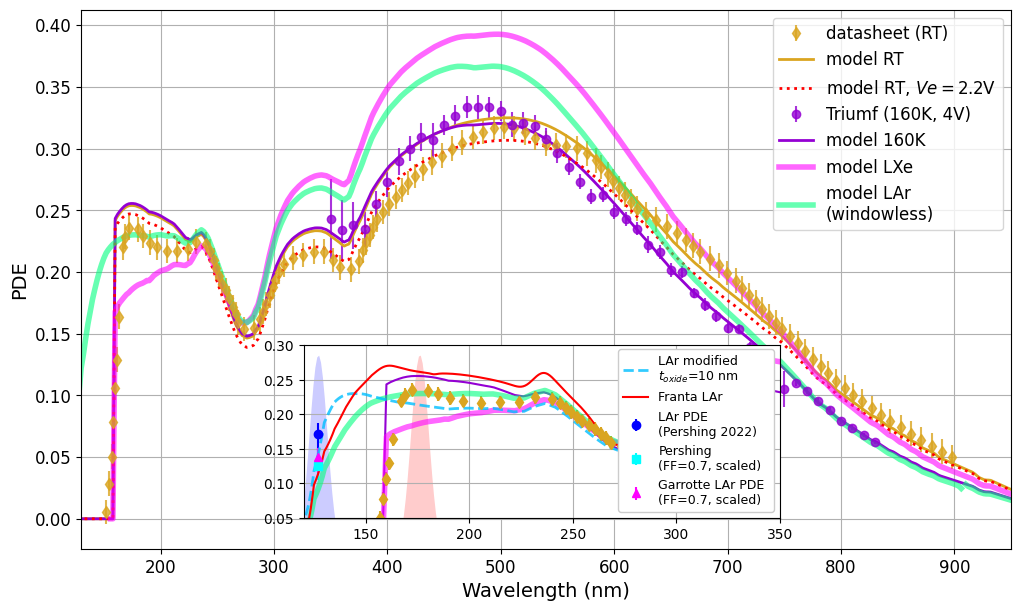}
\caption{\label{fig:hpk_datasheet}
Comparison of the HPK reported PDE (at room temperature - RT), measured PDE (TRIUMF cold) and the PDE model for different inputs. Increasing the model's temperature from 160~K to 273~K improves the Vis-NIR prediction compared to the datasheet. Thicker lines are the estimated PDE in LXe, and LAr without quartz window. Inset shows the VUV region in detail, including the Franta result (higher due to shalower $dp^*$ driven by curvature in $n, k$ data) and a second LAr prediction with a thinner oxide. The red dotted line is model with reduced $P_e$ ($V_e=2.2$~V), in better agreement with datasheet. The LAr datapoints are from \cite{Pershing_2022_argonPDE, garrote_2024_cryoPDE}, scaled by the $FF$ of each device.}
\end{figure}

The model is tested by extrapolating in wavelength and comparing with Hamamatsu reported vacuum PDE, and LXe measurements from literature. \autoref{fig:hpk_datasheet} shows the HPK reported PDE, the 4~V TRIUMF data and prediction at two temperatures. Beyond 500~nm, increasing the model temperature from 160~K to 273~K increases the PDE to approach the datasheet, giving confidence to the temperature dependence implementation above $\sim 500$~nm. Below 500~nm, the TRIUMF measurement and corresponding model extrapolation to the VUV is $\sim$2\% higher than the datasheet. Overall the model predicts the reported shape from 160-350~nm well. In the VUV, silicon quantum yield is applied to $P_e$ as outlined in \autoref{sec:internal_pde}\footnote{We approximate the quantum yield from \cite{Lewis_2025_qy} as $\eta(\lambda) = (\frac{300~\text{nm}}{\lambda})^{3/2}$}. Extrapolation to wavelengths relevant for ExCT are shown in inset of \autoref{fig:hpk_wlen_fit}.

%%%%%
%%%%%

The differences between data may be due to measurement systematics or temperature differences. More likely is device-to-device variation due to manufacturing and handling, which is difficult to account for analytically. We take this extrapolation as evidence for the predictive power of the model, showing VUV efficiency can be predicted using UV/IR efficiency, quantum yield, and VUV reflectivity to constrain oxide thickness.

The predicted LAr and LXe PDE show an increase in the visible region due to improved optical coupling. For LXe the transmission decreases below 250~nm as $n_{LXe} \approx n_{\sio} < n_{\si}$ and the enhancement from constructive interference is reduced. Reduction of UV LXe efficiency is consistent with other preliminary measurements \cite{Alex_2024LoLX2}. For the 175~nm centered xenon emission, the maximum achievable transmission in liquid is $\sim40\%$ with $\tox = 44$~nm. The VUV4's $\tox$ yields $\sim$55\% vacuum transmission and 38\% in liquid; the $\sim$17~nm oxide is well optimized LXe scintillation in gas or liquid. For FBK the average thickness yields transmission of 38\% in vacuum and 39\% in liquid, similar to the ray-optics values as oscillations are averaged over the emission spectrum.

The falloff in the PDE at 160~nm is due to absorption on the quartz window. In \cite{Pershing_2022_argonPDE} the window is removed and PDE for LAr scintillation of $\sim 14–17\%$ is reported for two VUV4 devices. Dividing by fill factor and assuming $iPDE=1$, the reported efficiencies require transmissions between $20-28\%$ at 127~nm. Our optical model cannot produce transmission this high, giving $T(128\text{nm})=17.0^{+1.4}_{-1.2}\% $ for the default dataset and $T(128\text{nm})=20.7^{+1.2}_{-0.6}\%$ with Franta. Within the constraints of our model, a combination of thinner oxide and reduced $k_{\sio}$ is required to achieve transmissions approaching 28\%. Perfect internal efficiency would be driven by an increase in charge collection efficiency (reduced $dp^*$) \cite{cce_cryogenic_1998, biroth_2015_cryogenic} or higher $P_e$ due to lower temperature.
% For example scaling $k_{\sio}$ by 50\% and reducing $\tox$ by 5, 3~nm yields 28\% transmission for default, Franta datasets respectively. 

\begin{figure}[htbp]
\centering
\includegraphics[width=0.8\linewidth]{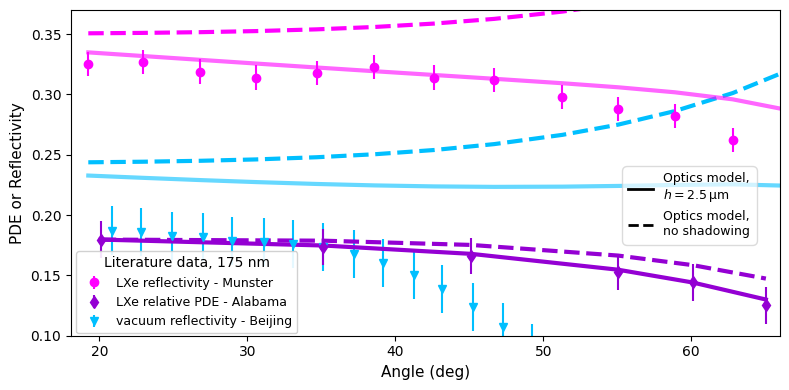}
\caption{Optical model compared to angular responses for VUV4 device for $175$~nm light or LXe emission. Solid lines show optical model including shadowing, the dashed lines without shadowing. Relative PDE data (Alabama, \cite{Nakarmi_2020}) is scaled vertically to compare with the model. In all cases shadowing better describes the data. Vacuum data (Beijing, \cite{sipmRef_nexo_guofu}) is lower than expectation from $R_{SiPM} = FF \cdot R_{\si}$, but predicted LXe reflectivity agree (Munster, \cite{Wagenpfeil_2021}).\label{fig:vuv4_reflectivity}.}
\end{figure}

In \autoref{fig:vuv4_reflectivity} we compare the model's angular predictions with shadowing to LXe measurements of reflectivity \cite{Wagenpfeil_2021} and relative PDE \cite{Nakarmi_2020}. Vacuum reflectivity from \cite{sipmRef_nexo_guofu} is also included showing the reduction relative to the assumption
$R(\theta){\mathrm{SiPM}} = FF(\theta)\cdot R{\mathrm{Si}}$. Our model predicts the absolute LXe reflectivity well (upper pink data and dashed line), and shadowing is clearly required to reproduce the monotonic decrease in reflectivity and best describes the relative PDE. 
\section{Discussion}
\label{sec:discussion}

%% FF(theta) dependance is probably tradeoff of FF factors listed near the top

%% for discussion

%% should include nested fits in Pe comparison plot
%% maybe 360 resonance is smeared out by dopants/impurities

% takeaway is assessment of Vbr between different methods is critical if enforcing Pe from independent measurement is goal
%% better parameterization

%% not sure how to organize % xx

%%%%%%%%%%%%%%%%%%%%%%%%%%%%%%%%%%%% -- %%%%%%%%%%%%%%%%%%%%%%%%%%%%%%%%%%%%
%% General fitting message
We have demonstrated that the many parameters of the PDE model can be constrained using combinations of relative optical measurements and absolute PDE. The model shows predictive power in extrapolating PDE to unmeasured wavelengths, angles, operational media and a subset of temperatures. Given the reliability of $n, k$ data in the visible-NIR and lack of structure, the extrapolation to dense media for modelling ExCT should be quite robust. The agreement between sequential and global fit methods indicate PDE for specific wavelengths may be modeled without full spectral information. % If only UV PDE is under study, $P_e(V)$ from PDE or rate-based measurements may be used; otherwise the PDE derived $P_e(V)$ is more accurate.
Including data below 350~nm should better constrain $dp^*$ regardless of fit method, and future work may investigate a `soft-edge' model with probabilistic charge collection between $0<z<dp^*$ which flattens the response about 360~nm.

%% data and liquid nobles
The model's underestimated LAr PDE emphasizes impact of oxide layer thickness and encourages further study of VUV optical constants at cryogenic temperatures. This is challenging as $\sio$ optical properties are known to vary with growth conditions and techniques \cite{sio2_properties1, sio2_properties2, SiO2_dep_2001_khodier}. Results from \cite{sio2_properties1} may be implemented in future work. The steep dispersion of liquid noble $n$ values near their scintillation response, and slope of \sio $n, k$ near 128~nm, underscores the importance of including spectral information in computation.%For LAr and LXe we have uncertainty in the exact value of $n$ at long wavelengths ($\lambda \sim 1000$~nm), and if the IR resonance used for the Sellmeier parameterization in \cite{lar_lxe_GRACE2017} is physical. In LXe the estimated range of values $n(1000\text{nm})=1.35-1.4$ only modifies the $\sim 80\%$ transmission into silicon by 1.2\%\footnote{1.35 taken from the pre-resonance region in \cite{lar_lxe_GRACE2017}, 1.4 from $n\approx \sqrt{\epsilon_r}$, $\epsilon_r=1.96$ reported in \cite{Aprile_2010}.}.

\subsection{Input Optical Datasets}
Although the Franta optical dataset can fit the PDE, results are rejected with limited confidence due to the poorer quality of fit and $P_h$ values higher than expectation from impact ionization ratios between electrons and holes\cite{gallina_avalanche_2019, mcintyre_avalanche_1973, oldham_avalanche}. The increased $P_h$ is not unexpected as the dataset describes ultrapure silicon, while SiPMs are heavily doped which increases photoabsorption \cite{photoAbs_doping_Abroug2008}. These results illustrates some quasi-degeneracy between photoabsorption data, junction size and $P_h$ in the NIR; PDE data extending to 1050~nm would help resolve this. Alternatively, direct study of $P_h$ could be achieved using a two-photon absorption technique \cite{Pape_2024} or by studying the rate of delayed crosstalk, which is expected to be entirely hole-driven in P-on-N devices. Given a known doping profile, implementation of the complete Franta optical model, which includes the impact of dopants on photoabsorption, would be possible.

\subsection{Angular Measurements}
We found the shadowing effect on the HPK-VUV4 essential to reproduce other angular measurements at $\sim 175$~nm (\autoref{fig:vuv4_reflectivity}). No shadowing was observed for the FBK device likely due to smaller or asymmetric resistors. At longer wavelengths the effective $FF(\lambda)$ can increase as the charge collection region is not uniform in depth\cite{ACERBI2018_FF}. We suspect the interplay between this effect and shadowing drives the measured increase of $FF(\theta)$ with wavelength.

For the FBK angular scans, the variation in oscillations (and oxide thickness) may be explained by surface thickness variations, as the beam's exact position varied between installs. Regardless, discrepancies in the FBK data highlighted the need for a more precise understanding of the experimental setup's optics, namely effects of the lamp's underlying spectrum and MC transfer function. As oscillations in equation \autoref{eq:transmission} are not symmetric with respect to $\theta$ or $\lambda$, the averaged transmission's maxima can shift depending on transfer function shape.

%% atp cmparison section
\subsection{Comparing Avalanche Triggering Probabilities} % xx
\label{sec:ATP}

\begin{figure}[htbp]
\centering
\includegraphics[width=.9\linewidth]{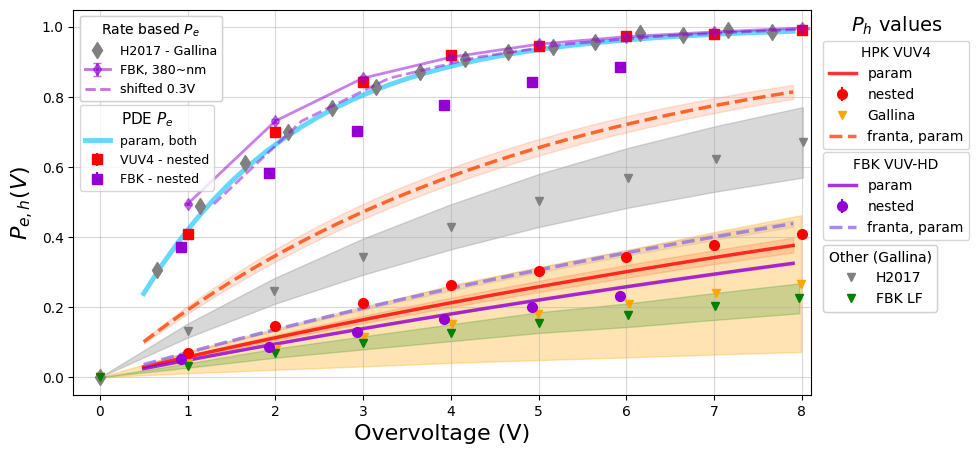}
\caption{\label{fig:Ph}
Comparison of avalanche probabilities. Upper curves are $P_e$ from fits or rate-based measurement. All parameterized fits give essentially the same $P_e$ curve (gray line). The unparameterized or `nested' $P_e$ values deviate slightly from the parameterization. Lower curves are $P_h$ values, taken from this work or \cite{gallina_avalanche_2019}. The Franta driven $P_h$ values are higher than the default dataset's due to the lower photoabsorption in the NIR and are less supported by other literature and physical expectation.}
\end{figure}

The nested $P_e$ fits perform best and are less susceptible to bias from low-voltage data or inaccurate breakdown voltages. The exact mapping between rate-based or PDE driven $P_e(V)$ (3A vs 3B in \autoref{tab:fitting_procedure}) is difficult, requiring the absolute accuracy of UV transmission and $FF$. Additionally, the assumption of a sharp transition from $P_e$ to $P_h$ about $X_{PN}$ can allow longer wavelengths to bias PDE driven $P_e$. Fitting for the $V_e$ shape parameter is sensitive to the breakdown voltage, which can differ between IV or gain based analysis; our rate based measurements ($V_e = 1.5-1.8 \pm 0.1$) agree with PDE driven $V_e$ only if a shift in voltage is applied, which can be seen in the shifted dotted lines of \autoref{fig:Ph}. We suspect the low nested FBK $P_e$ values (purple squares in \autoref{fig:Ph}) are compensation for other overestimated terms; decreasing $FF=0.8\to0.7$ drives $P_e$ towards unity as expected.

The takeaway is that $PDE$ derived $P_e(V)$ should be used when calculating the absolute efficiency over many wavelengths. If studying voltage dependence in a narrow UV regime, especially in a relative measurement, either $P_e(V)$ value may be used. A $P_e$ parameterization function with two degrees of freedom may reduce the sensitivity to breakdown voltage or equivalently better describe lower voltage PDE. % For HPK, the shape of $PDE(V)$ and $P_e(V)$ curves also differ, likely due to a systematic difference between the $Q_A(V)$ correction and primary pulse identification method.

\autoref{fig:Ph} compares our various $P_h$, $P_e$ curves and those reported by Gallina \cite{gallina_avalanche_2019}. The similarity of the $P_e$ curves is remarkable, despite coming from different devices, optical data, measurement techniques and temperatures. The assumption that $P_e$ saturates to unity was validated for HPK. Further work should study the universality of $P_e$ between devices at temperatures below 160~K, and shape difference at low overvoltages. % The rate-based $P_e$ measurement is horizontally offset from the fit values. Allowing for a horizontal shift in voltage of 0.1~V, owing to uncertainty in breakdown voltage calculation between datasets, the curves are similar.
The $P_h$ curves have larger variance. The default-dataset derived $P_h$ values are low and fall within the large uncertainty bands of the VUV4 device from \cite{gallina_avalanche_2019}. Higher Franta-derived values are included, with the resulting VUV4 parameterization ($V_e = 4.7$) exceeding that of the H2017 device in \cite{gallina_avalanche_2019}. %$P_h$ between fits are in better agreement for the FBK device likely due to stronger constraint on $\tox$ from the oscillations.
A larger variance in $P_h$ across different devices is not unexpected at moderate overvoltages.%, as the lower mobility of holes prevents $P_h$ from quickly saturating.
The sharp transition at $X_{PN}$ assumption may contribute, as larger junctions can have a less-well defined singular $P_h$ value.
%This may explain the underestimation of the VUV4's PDE near 480~nm (where both $W_p, W_n > 0$).

\subsection{Temperature Dependence} % xx
Preliminary $P_e(V)$ measurements showed negligible temperature dependence between 160~K and 253~K. The PDE reported in \cite{gallina_performance_2022} at both 160~K and 300~K, once quantum yield is included, are similar to those measured here. The shape of LAr $PDE(V)$ from Pershing \cite{Pershing_2022_argonPDE} actually suggest lower $P_e$ at 90~K, assuming the quantum yield in \cite{Lewis_2025_qy} is monotonic in energy. Lower $P_e$ with temperature is possible due to carrier freezeout overcoming the increase in mobility, or measurement contamination by fluorescence, which would contribute $P_h$ and inflate reported PDE.

The third FBK angular campaign measured relative $PDE(\theta)$ at 213~K and 163~K, producing identical oscillation patterns at both temperatures. This suggests the surface optics, at least at 400 and 600~nm, are unperturbed by those temperatures. The parameterized temperature dependence of the photo-absorption appears reliable, although in the future we hope validate it on PDE data below the 440~nm limit from \cite{siBandgap_dmPaper}. Future work may include the effect of carrier freezeout; the complete temperature response of PDE below 100~K is complex and requires further modeling.

\section{Optimizing PDE for Liquid Nobles and Quantum Sensing}
\label{sec:maximize}

We estimate the theoretical upper limit of PDE for P-on-N devices using the single $\sio$ layer surface model. Maximizing PDE requires large fill factor $FF$, transmission, and junction geometries favoring electron-initiated avalanches ($W_p \sim 1$), assuming operation with $P_e(V) \sim 1$ while correlated avalanches remain tractable. The PDE becomes: $PDE_{\text{max}} \approx FF \cdot T_{\text{max}} \cdot W_p$.

Achieving high $FF$ remains fabrication-limited due to maintaining a low boundary field within each SPAD, quenching resistors, and isolation trenches. Backside-illuminated structures show promise for $FF \sim 100\%$ \cite{ninkovic_avalanche_2007}. Multilayer anti-reflective coatings enable near-unity transmission in the visible and improved UV performance \cite{uvMulti_Hamden2011, JIA_solarcells_broadband_2017, solar_multilayer_review, multilayer_2005}, though at increased cost and complexity outside this model's scope. \autoref{tab:quantum} lists optimized parameters for particle physics applications and qubit control, free-space quantum communication \cite{liao_satellite--ground_2017}. Oxide thicknesses maximize normal-incidence transmission, while junction geometries ensure $\sim90\%$ absorption.
\begin{table}[ht!]
\centering
\caption{\label{tab:quantum} Optimized $\sio$ thickness and junction geometries for maximizing PDE. Relevant geometries achieve $W_p=0.9$; for VUV wavelengths $dp^*$ is given and junction center $X_{PN}$ for others.}
\smallskip
\renewcommand{\arraystretch}{1.3}  % Default is 1.0
{\footnotesize
\begin{tabular}{|c|c|c|c|c|}
\hline
\textbf{Application} & \textbf{{$\bm{\lambda}$ (nm)}} & \makecell{\textbf{Maximum}\\\textbf{transmission}} & \textbf{$\bm{\tox}$ (nm)} & \textbf{$\bm{dp^*}$ or $\bm{X_{PN}}$} \\
\Xhline{3\arrayrulewidth}
LAr in vacuum & 127 $\pm$ 10 & 0.44 & 4.7 & 0.6~nm \\ \hline
LAr           & 127 $\pm$ 10 & 0.38 & 3.45 & 0.6~nm \\ \hline
Xe-doped LAr  & 175 $\pm$ 10 & 0.46 & 17.3 & 0.5~nm \\ \hline
LXe           & 175 $\pm$ 10 & 0.40 & 44 & 0.5~nm \\ \hline
Xe in $\ce{CF_4}$ & 175 $\pm$ 10 & 0.47 & 17.2 & 0.5~nm \\ \hline
TPB \cite{TPB_Francini_2013}, PEN \cite{PEN_Kuniak2019} in LAr & 430 $\pm$ 80 & 0.76 & 71 & \small{$X_{PN}=$~550~nm} \\
\Xhline{3\arrayrulewidth}
Diamond NV center \cite{childress_diamond_2013} & 532 & 0.89 & 91.3 & 1.7 $\um$ \\ \hline
Sodium D-line & 589 & 0.91 & 101  & 3.3 $\um$ \\ \hline
NV center (zero-phonon) \cite{NV_zerophonon} & 637 & 0.91 & 328  & 5.6 $\um$ \\ \hline
Rubidium D1, D2 lines \cite{mottola_optical_2023} & 780, 795 & 0.93 & 410, 671 & 21, 27 $\um$ \\ \hline
Satellite-Ground QKD \cite{liao_satellite--ground_2017} & 850 & 0.93 & 731  & 45 $\um$ \\
\hline
\end{tabular}
} % end of font size
\end{table}

For LXe scintillation both devices studied achieve $iPDE \sim 90\%$. In LAr, PDE approaching $\sim35\%$ is possible for very thin oxides. Xe-doped LAr can achieve $\sim6\%$ higher PDE than pure LXe by via constructive interference. In $\ce{CF4}$, the operational medium of SBC\cite{sbc_snowmass2021}, the transmission of xenon light is noticeably higher than in vacuum\footnote{$n=1.24$ is taken for liquid $\ce{CF4}$ from \cite{index_cf4_chemPhys, index_cf4_french}}. In quantum-sensing, transmission near $90\%$ combined with $W_p \approx 0.9$ implies PDE approaching $80\%$, given $FF \sim 100\%$. Digital SiPMs may provide the best path forward due to insensitivity to correlated avalanches \cite{Pratte_2021}. This analysis ignores the complexity of doping and engineering suitable electric-field profiles.
\section{Conclusion}

We have demonstrated the application of our broadband, multivariable PDE model on VUV sensitive devices. The PDE data for the Hamamatsu VUV4 and FBK VUV-HD technology was fit, yielding junction parameters, avalanche probabilities and oxide thicknesses allowing PDE prediction, validated against other reported measurements. Compared to the HPK-VUV4, the FBK VUV-HD's oxide is roughly 100 times thicker and the junction almost half the size. All measured $P_e$ values are similar, with a large spread in $P_h$ values. For operation in liquid nobles, we consider the PDE extrapolation at long wavelengths relevant to ExCT quite robust. Further model investigation and validation in the VUV, especially for LAr emission, is required. Immediate opportunities for expanding the model are measuring $P_e(V)$ and relative UV $PDE(\theta)$ over a wide temperature range. This model provides a foundation for improved detector simulation and SiPM characterization and understanding, and we hope combinations of the methods outlined here can be leveraged by other groups to characterize device PDE and extrapolate when measurement in the region of interest is not possible.  

The PDE model and lookup tables for simulation will be made available online.

%We anticipate future designs can push $FF$ close to unity while retaining high avalanche efficiency and timing resolution.

\section*{Acknowledgments}
The authors gratefully acknowledge support from Canadian Foundation for Innovation Fund (CFI) 2017. Additional support was provided by a grant from Canada Natural Sciences and Engineering Research Council of Canada (NSERC). We thank the Canadian taxpayers for supporting physics research.

%% For citations use: 
%%       \citet{<label>} ==> Lamport [21]
%%       \citep{<label>} ==> [21]
%%       \cite{<label>} ==> [21]
%%
\bibliographystyle{elsarticle-num} 
\bibliography{
    zBib-opticalData,
    zBib-SiPMs,
    zBib-yield_siliconAndDevices,
    zBib-siliconProperties,
    zBib-nonSilicon,
    zBib-physics_applications,
    zBib-Quantum_and_FSO,
    zBib-BSI_and_digital
}

@article{ninkovic_avalanche_2007,
	series = {Imaging 2006},
	title = {The avalanche drift diode—{A} back illumination drift silicon photomultiplier},
	volume = {580},
	issn = {0168-9002},
	url = {https://www.sciencedirect.com/science/article/pii/S0168900207013137},
	doi = {10.1016/j.nima.2007.06.060},
	number = {2},
	urldate = {2025-05-28},
	journal = {Nuclear Instruments and Methods in Physics Research Section A: Accelerators, Spectrometers, Detectors and Associated Equipment},
	author = {Ninković, Jelena and Eckhart, Rouven and Hartmann, Robert and Holl, Peter and Koitsch, Christian and Lutz, Gerhard and Merck, Christine and Mirzoyan, Razmik and Moser, Hans-Günther and Otte, Adam-Nepomuk and Richter, Rainer and Schaller, Gerhard and Schopper, Florian and Soltau, Heike and Teshima, Masahiro and Vâlceanu, George},
	month = oct,
	year = {2007},
	pages = {1013--1015},
}

@Article{Pratte_2021,
AUTHOR = {Pratte, Jean-François and Nolet, Frédéric and Parent, Samuel and Vachon, Frédéric and Roy, Nicolas and Rossignol, Tommy and Deslandes, Keven and Dautet, Henri and Fontaine, Réjean and Charlebois, Serge A.},
TITLE = {3D Photon-To-Digital Converter for Radiation Instrumentation: Motivation and Future Works},
JOURNAL = {Sensors},
VOLUME = {21},
YEAR = {2021},
NUMBER = {2},
ARTICLE-NUMBER = {598},
URL = {https://www.mdpi.com/1424-8220/21/2/598},
PubMedID = {33467016},
ISSN = {1424-8220},
DOI = {10.3390/s21020598}
}

@article{liao_satellite--ground_2017,
	title = {Satellite-to-ground quantum key distribution},
	volume = {549},
	copyright = {2017 Macmillan Publishers Limited, part of Springer Nature. All rights reserved.},
	issn = {1476-4687},
	url = {https://www.nature.com/articles/nature23655},
	doi = {10.1038/nature23655},
	language = {en},
	number = {7670},
	urldate = {2024-02-15},
	journal = {Nature},
	author = {Liao, Sheng-Kai and Cai, Wen-Qi and Liu, Wei-Yue and Zhang, Liang and Li, Yang and Ren, Ji-Gang and Yin, Juan and Shen, Qi and Cao, Yuan and Li, Zheng-Ping and Li, Feng-Zhi and Chen, Xia-Wei and Sun, Li-Hua and Jia, Jian-Jun and Wu, Jin-Cai and Jiang, Xiao-Jun and Wang, Jian-Feng and Huang, Yong-Mei and Wang, Qiang and Zhou, Yi-Lin and Deng, Lei and Xi, Tao and Ma, Lu and Hu, Tai and Zhang, Qiang and Chen, Yu-Ao and Liu, Nai-Le and Wang, Xiang-Bin and Zhu, Zhen-Cai and Lu, Chao-Yang and Shu, Rong and Peng, Cheng-Zhi and Wang, Jian-Yu and Pan, Jian-Wei},
	month = sep,
	year = {2017},
	note = {Number: 7670
Publisher: Nature Publishing Group},
	pages = {43--47},
}

@article{hadfield_single-photon_2009,
	title = {Single-photon detectors for optical quantum information applications},
	volume = {3},
	copyright = {2009 Springer Nature Limited},
	issn = {1749-4893},
	url = {https://www.nature.com/articles/nphoton.2009.230},
	doi = {10.1038/nphoton.2009.230},
	language = {en},
	number = {12},
	urldate = {2025-06-24},
	journal = {Nature Photonics},
	author = {Hadfield, Robert H.},
	month = dec,
	year = {2009},
	note = {Publisher: Nature Publishing Group},
	pages = {696--705},
}

@inproceedings{ara_shawkat_single_2023,
	title = {Single {Photon} {Detectors} for {Quantum} {Computing}},
	url = {https://ieeexplore.ieee.org/document/10130206},
	doi = {10.1109/DCAS57389.2023.10130206},
	urldate = {2025-06-24},
	booktitle = {2023 {IEEE} 16th {Dallas} {Circuits} and {Systems} {Conference} ({DCAS})},
	author = {Ara Shawkat, Mst Shamim and Hasan, Sajid and McFarlane, Nicole},
	month = apr,
	year = {2023},
	pages = {1--4},
}

@article{childress_diamond_2013,
	title = {Diamond {NV} centers for quantum computing and quantum networks},
	volume = {38},
	issn = {0883-7694, 1938-1425},
	url = {https://www.cambridge.org/core/journals/mrs-bulletin/article/diamond-nv-centers-for-quantum-computing-and-quantum-networks/978A4B94242CF28F9C60F0D9E95E9CBD},
	doi = {10.1557/mrs.2013.20},
	language = {en},
	number = {2},
	urldate = {2025-06-24},
	journal = {MRS Bulletin},
	author = {Childress, Lilian and Hanson, Ronald},
	month = feb,
	year = {2013},
	pages = {134--138},
}

@article{mottola_optical_2023,
	title = {Optical {Memory} in a {Microfabricated} {Rubidium} {Vapor} {Cell}},
	volume = {131},
	url = {https://link.aps.org/doi/10.1103/PhysRevLett.131.260801},
	doi = {10.1103/PhysRevLett.131.260801},
	number = {26},
	urldate = {2025-06-24},
	journal = {Physical Review Letters},
	author = {Mottola, Roberto and Buser, Gianni and Treutlein, Philipp},
	month = dec,
	year = {2023},
	note = {Publisher: American Physical Society},
	pages = {260801},
}

@article{ACERBI2018_FF,
title = {Silicon photomultipliers and single-photon avalanche diodes with enhanced NIR detection efficiency at FBK},
journal = {Nuclear Instruments and Methods in Physics Research Section A: Accelerators, Spectrometers, Detectors and Associated Equipment},
volume = {912},
pages = {309-314},
year = {2018},
note = {New Developments In Photodetection 2017},
issn = {0168-9002},
doi = {https://doi.org/10.1016/j.nima.2017.11.098},
url = {https://www.sciencedirect.com/science/article/pii/S0168900217313542},
author = {Fabio Acerbi and Giovanni Paternoster and Alberto Gola and Nicola Zorzi and Claudio Piemonte}
}

@ARTICLE{sipmRef_nexo_guofu,
  author={Lv, P. and Cao, G. F. and Wen, L. J. and others},
  journal={IEEE Transactions on Nuclear Science}, 
  title={Reflectance of Silicon Photomultipliers at Vacuum Ultraviolet Wavelengths}, 
  year={2020},
  volume={67},
  number={12},
  pages={2501-2510},
  doi={10.1109/TNS.2020.3035172}}

@article{Wagenpfeil_2021,
doi = {10.1088/1748-0221/16/08/P08002},
url = {https://doi.org/10.1088/1748-0221/16/08/P08002},
year = {2021},
month = {aug},
publisher = {IOP Publishing},
volume = {16},
number = {08},
pages = {P08002},
author = {Wagenpfeil, M. and Ziegler, T. and Schneider, J. and Fieguth, A. and Murra, M. and Schulte, D. and Althueser, L. and Huhmann, C. and Weinheimer, C. and Michel, T. and others},
title = {Reflectivity of VUV-sensitive silicon photomultipliers in liquid Xenon},
journal = {Journal of Instrumentation}
}

@article{Nakarmi_2020,
doi = {10.1088/1748-0221/15/01/P01019},
url = {https://doi.org/10.1088/1748-0221/15/01/P01019},
year = {2020},
month = {jan},
publisher = {},
volume = {15},
number = {01},
pages = {P01019},
author = {Nakarmi, P. and Ostrovskiy, I. and Soma, A.K. and Retière, F. and others},
title = {Reflectivity and PDE of VUV4 Hamamatsu SiPMs in liquid xenon},
journal = {Journal of Instrumentation}
}

@misc{lolx_ext_2025,
      title={Characterization of external cross-talk from silicon photomultipliers in a liquid xenon detector}, 
      author={D Gallacher and A. de St. Croix and S. Bron and B. M. Rebeiro and T. McElroy and S. Al Kharusi and T. Brunner and C. Chambers and B. Chana and Z. Charlesworth and E. Egan and M. Francesconi and L. Galli and P. Giampa and D. Goeldi and S. Lavoie and J. Lefebvre and X. Li and C. Malbrunot and P. Margetak and N. Massacret and S. C. Nowicki and H. Rasiwala and K. Raymond and F. Retière and S. Rottoo and L. Rudolph and M. A. Tétrault and S. Viel and N. V. H. Viet and L. Xie},
      year={2025},
      eprint={2502.15991},
      archivePrefix={arXiv},
      primaryClass={physics.ins-det},
      url={https://arxiv.org/abs/2502.15991} 
}

@mastersthesis{kurtis_thesis,
  author       = {Kurtis Raymond},
  title        = {Stimulated secondary emission from single photon avalanche diodes},
  school       = {Simon Fraser University},
  year         = {2024},
  address      = {Burnaby, BC, Canada},
  month        = {May},
  note         = {Master's Thesis},
}

@article{raymond_stimulated_TED_2024,
	title = {Stimulated {Secondary} {Emission} of {Single}-{Photon} {Avalanche} {Diodes}},
	issn = {1557-9646},
	url = {https://ieeexplore.ieee.org/document/10713286},
	doi = {10.1109/TED.2024.3469918},
	urldate = {2024-10-15},
	journal = {IEEE Transactions on Electron Devices},
	author = {Raymond, Kurtis and Retière, Fabrice and Lewis, Harry and Capra, Andrea and McCarthy, Duncan and Croix, Austin de St and Gallina, Giacomo and McLaughlin, Joe and Martin, Juliette and Massacret, Nicolas and Agnes, Paolo and Underwood, Ryan and Koulosousas, Seraphim and Margetak, Peter},
	year = {2024},
	note = {Conference Name: IEEE Transactions on Electron Devices},
	pages = {1--9}
}

@article{joe_emission_2021,
AUTHOR = {McLaughlin, Joseph Biagio and Gallina, Giacomo and Retière, Fabrice and De St. Croix, Austin and Giampa, Pietro and Mahtab, Mahsa and Margetak, Peter and Martin, Lars and Massacret, Nicolas and Monroe, Jocelyn and Patel, Mayur and Raymond, Kurtis and Roiseux, Jolie and Xie, Liang and Zhang, Guoqing},
TITLE = {Characterisation of SiPM Photon Emission in the Dark},
JOURNAL = {Sensors},
VOLUME = {21},
YEAR = {2021},
NUMBER = {17},
ARTICLE-NUMBER = {5947},
URL = {https://www.mdpi.com/1424-8220/21/17/5947},
PubMedID = {34502838},
ISSN = {1424-8220},
DOI = {10.3390/s21175947}
}

@article{Gibbons_2024_flashlight,
doi = {10.1088/1748-0221/19/01/P01013},
url = {https://doi.org/10.1088/1748-0221/19/01/P01013},
year = {2024},
month = {jan},
publisher = {IOP Publishing},
volume = {19},
number = {01},
pages = {P01013},
author = {Gibbons, R. and Chen, H. and Haselschwardt, S.J. and Xia, Q. and Sorensen, P.},
title = {Why would you put a flashlight in a dark matter detector?},
journal = {Journal of Instrumentation}
}

@article{hampel_optical_2020,
	title = {Optical crosstalk in {SiPMs}},
	volume = {976},
	issn = {0168-9002},
	url = {https://www.sciencedirect.com/science/article/pii/S0168900220306586},
	doi = {10.1016/j.nima.2020.164262},
	urldate = {2023-10-12},
	journal = {Nuclear Instruments and Methods in Physics Research Section A: Accelerators, Spectrometers, Detectors and Associated Equipment},
	author = {Hampel, M. R. and Fuster, A. and Varela, C. and Platino, M. and Almela, A. and Lucero, A. and Wundheiler, B. and Etchegoyen, A.},
	month = oct,
	year = {2020},
	pages = {164262},
}

@article{acerbi_2023,
title = {NUV and VUV sensitive Silicon Photomultipliers technologies optimized for operation at cryogenic temperatures},
journal = {Nuclear Instruments and Methods in Physics Research Section A: Accelerators, Spectrometers, Detectors and Associated Equipment},
volume = {1046},
pages = {167683},
year = {2023},
issn = {0168-9002},
doi = {https://doi.org/10.1016/j.nima.2022.167683},
url = {https://www.sciencedirect.com/science/article/pii/S0168900222009755},
author = {Fabio Acerbi and Giovanni Paternoster and Stefano Merzi and Nicola Zorzi and Alberto Gola}
}

@article{sipm_temp_pde_collazuol_2011,
title = {Studies of silicon photomultipliers at cryogenic temperatures},
journal = {Nuclear Instruments and Methods in Physics Research Section A: Accelerators, Spectrometers, Detectors and Associated Equipment},
volume = {628},
number = {1},
pages = {389-392},
year = {2011},
note = {VCI 2010},
issn = {0168-9002},
doi = {https://doi.org/10.1016/j.nima.2010.07.008},
url = {https://www.sciencedirect.com/science/article/pii/S0168900210015500},
author = {G. Collazuol and M.G. Bisogni and S. Marcatili and C. Piemonte and A. {Del Guerra}}
}

@article{biroth_2015_cryogenic,
title = {Silicon photomultiplier properties at cryogenic temperatures},
journal = {Nuclear Instruments and Methods in Physics Research Section A: Accelerators, Spectrometers, Detectors and Associated Equipment},
volume = {787},
pages = {68-71},
year = {2015},
note = {New Developments in Photodetection NDIP14},
issn = {0168-9002},
doi = {https://doi.org/10.1016/j.nima.2014.11.020},
url = {https://www.sciencedirect.com/science/article/pii/S0168900214012844},
author = {M. Biroth and P. Achenbach and E. Downie and A. Thomas}
}

@article{garrote_2024_cryoPDE,
title = {Measurement of the photon detection efficiency of Hamamatsu VUV4 SiPMS at cryogenic temperature},
journal = {Nuclear Instruments and Methods in Physics Research Section A: Accelerators, Spectrometers, Detectors and Associated Equipment},
volume = {1064},
pages = {169347},
year = {2024},
issn = {0168-9002},
doi = {https://doi.org/10.1016/j.nima.2024.169347},
url = {https://www.sciencedirect.com/science/article/pii/S0168900224002730},
author = {R. Álvarez-Garrote and E. Calvo and A. Canto and J.I. Crespo-Anadón and C. Cuesta and A. {de la Torre Rojo} and I. Gil-Botella and S. {Manthey Corchado} and I. Martín and C. Palomares and L. Pérez-Molina and A. {Verdugo de Osa}}
}

@article{cce_cryogenic_1998,
title = {Evidence for charge collection efficiency recovery in heavily irradiated silicon detectors operated at cryogenic temperatures},
journal = {Nuclear Instruments and Methods in Physics Research Section A: Accelerators, Spectrometers, Detectors and Associated Equipment},
volume = {413},
number = {2},
pages = {475-478},
year = {1998},
issn = {0168-9002},
doi = {https://doi.org/10.1016/S0168-9002(98)00673-1},
url = {https://www.sciencedirect.com/science/article/pii/S0168900298006731},
author = {Vittorio Giulio Palmieri and Kurt Borer and Stefan Janos and Cinzia {Da Viá} and  {Luca Casagrande}}
}

@misc{cryogenic_2025_rivera,
      title={Cryogenic operation of silicon photomultiplier arrays}, 
      author={Esteban Currás-Rivera and Frédéric Blanc and Guido Haefeli and Radoslav Marchevski and Federico Ronchetti and Olivier Schneider and Lesya Shchutska and Gianluca Zunica},
      year={2025},
      eprint={2502.02235},
      archivePrefix={arXiv},
      primaryClass={hep-ex},
      url={https://arxiv.org/abs/2502.02235} 
}

@article{Wang_2021_cryo_sCurve,
doi = {10.1088/1748-0221/16/07/P07021},
url = {https://doi.org/10.1088/1748-0221/16/07/P07021},
year = {2021},
month = {jul},
publisher = {IOP Publishing},
volume = {16},
number = {07},
pages = {P07021},
author = {Wang, L. and Guan, M.Y. and Qin, H.J. and Guo, C. and Sun, X.L. and Yang, C.G. and Zhao, Q. and Liu, J.C. and Zhang, P. and Zhang, Y.P. and Xiong, W.X. and Wei, Y.T. and Gan, Y.Y. and Li, J.J.},
title = {Characterization of VUV4 SiPM for liquid argon detector},
journal = {Journal of Instrumentation}
}

@article{Borden_pdeTemp_2024,
doi = {10.1088/1748-0221/19/12/P12014},
url = {https://doi.org/10.1088/1748-0221/19/12/P12014},
year = {2024},
month = {dec},
publisher = {IOP Publishing},
volume = {19},
number = {12},
pages = {P12014},
author = {Borden, S. and Detwiler, J.A. and Pettus, W. and Ruof, N.W.},
title = {Characterization of Silicon Photomultiplier Photon Detection Efficiency at Liquid Nitrogen Temperature},
journal = {Journal of Instrumentation}
}

@misc{gu_2025_heliumPDE,
      title={Characterization of FBK NUV-HD-Cryo SiPMs near LHe temperature}, 
      author={Fengbo Gu and Junhui Liao and Jiangfeng Zhou and Meiyuenan Ma and Yuanning Gao and Zhaohua Peng and Jian Zheng and Guangpeng An and Lifeng Zhang and Lei Zhang and Zhuo Liang and Xiuliang Zhao and Fabio Acerbi and Andrea Ficorella and Alberto Gola and Laura Parellada Monreal},
      year={2025},
      eprint={2311.10497},
      archivePrefix={arXiv},
      primaryClass={physics.ins-det},
      url={https://arxiv.org/abs/2311.10497}
}

@article{PKLightfoot_2008,
doi = {10.1088/1748-0221/3/10/P10001},
url = {https://doi.org/10.1088/1748-0221/3/10/P10001},
year = {2008},
month = {oct},
publisher = {},
volume = {3},
number = {10},
pages = {P10001},
author = {P K Lightfoot and G J Barker and K Mavrokoridis and Y A Ramachers and N J C Spooner},
title = {Characterisation of a silicon photomultiplier device for applications 
in liquid argon based neutrino physics and dark matter searches},
journal = {Journal of Instrumentation}
}

@ARTICLE{gallina_avalanche_2019,
  author={Gallina, G. and Retière, F. and Giampa, P. and Kroeger, J. and Margetak, P. and Byrne Mamahit, S. and De St. Croix, A. and Edaltafar, F. and Martin, L. and Massacret, N. and Ward, M. and Zhang, G.},
  journal={IEEE Transactions on Electron Devices}, 
  title={Characterization of SiPM Avalanche Triggering Probabilities}, 
  year={2019},
  volume={66},
  number={10},
  pages={4228-4234},
  doi={10.1109/TED.2019.2935690}
}

@ARTICLE{biroth_icasipm_2018,
  author= {Biroth, M. and Achenbach and P. Lauth W. and Thomas, A.},
  title = {An analytical approach to predict fundamental cryogenic properties of silicon photomultipliers},
  year = {2018},
  address = {Palais Hirsch, Schwetzingen},
  note = {Conference Presentation},
  journal = {ICASIPM},
  url={https://indico.gsi.de/event/6990/contributions/31517/}
}

@article{Zappala_2016_efficiency,
doi = {10.1088/1748-0221/11/11/P11010},
url = {https://doi.org/10.1088/1748-0221/11/11/P11010},
year = {2016},
month = {nov},
publisher = {},
volume = {11},
number = {11},
pages = {P11010},
author = {Zappalà, G. and Acerbi, F. and Ferri, A. and Gola, A. and Paternoster, G. and Regazzoni, V. and Zorzi, N. and Piemonte, C.},
title = {Study of the photo-detection efficiency of FBK High-Density silicon photomultipliers},
journal = {Journal of Instrumentation}
}

@article{Pershing_2022_argonPDE,
   title={Performance of Hamamatsu VUV4 SiPMs for detecting liquid argon scintillation},
   volume={17},
   ISSN={1748-0221},
   url={https://iopscience.iop.org/article/10.1088/1748-0221/17/04/P04017},
   DOI={10.1088/1748-0221/17/04/p04017},
   number={04},
   journal={Journal of Instrumentation},
   publisher={IOP Publishing},
   author={Pershing, T. and Xu, J. and Bernard, E. and Kingston, J. and Mizrachi, E. and Brodsky, J. and Razeto, A. and Kachru, P. and Bernstein, A. and Pantic, E. and Jovanovic, I.},
   year={2022},
   month=apr, pages={P04017}
}

@article{butcher_method_2017,
	title = {A method for characterizing after-pulsing and dark noise of {PMTs} and {SiPMs}},
	volume = {875},
	doi = {10.1016/j.nima.2017.08.035},
	journal = {Nuclear Instruments and Methods in Physics Research Section A: Accelerators, Spectrometers, Detectors and Associated Equipment},
	author = {Butcher, Alistair and Doria, Luca and Monroe, Jeffrey and Retiere, Fabrice and Smith, Brianna and Walding, Joseph},
	month = mar,
	year = {2017}
}

@article{gallina_performance_2022,
	title = {Performance of novel {VUV}-sensitive {Silicon} {Photo}-{Multipliers} for {nEXO}},
	volume = {82},
	issn = {1434-6052},
	url = {https://www.osti.gov/pages/biblio/1903455},
	doi = {10.1140/epjc/s10052-022-11072-8},
	language = {English},
	number = {12},
	urldate = {2023-10-12},
	journal = {European Physical Journal. C, Particles and Fields (Online)},
	author = {Gallina, G. and Guan, Y. and et al.},
	month = dec,
	year = {2022},
	note = {Institution: Stanford Univ., CA (United States); Pacific Northwest National Laboratory (PNNL), Richland, WA (United States); Oak Ridge National Laboratory (ORNL), Oak Ridge, TN (United States)
Number: PNNL-SA-179827
Publisher: Springer Nature},
	}

@article{gallina_characterization_2019-1,
	title = {Characterization of the {Hamamatsu} {VUV4} {MPPCs} for {nEXO}},
	volume = {940},
	issn = {0168-9002},
	url = {https://www.sciencedirect.com/science/article/pii/S0168900219308034},
	doi = {10.1016/j.nima.2019.05.096},
	urldate = {2023-10-17},
	journal = {Nuclear Instruments and Methods in Physics Research Section A: Accelerators, Spectrometers, Detectors and Associated Equipment},
	author = {Gallina, G. and Giampa, P. and et al.},
	month = oct,
	year = {2019},
	pages = {371--379},
}

@article{baudis_characterisation_2018,
	title = {Characterisation of {Silicon} {Photomultipliers} for liquid xenon detectors},
	volume = {13},
	issn = {1748-0221},
	url = {https://dx.doi.org/10.1088/1748-0221/13/10/P10022},
	doi = {10.1088/1748-0221/13/10/P10022},
	language = {en},
	number = {10},
	urldate = {2024-02-21},
	journal = {Journal of Instrumentation},
	author = {Baudis, L. and Galloway, M. and Kish, A. and Marentini, C. and Wulf, J.},
	month = oct,
	year = {2018},
	pages = {P10022},
}

@article{nSi_aspnes,
  title = {Dielectric functions and optical parameters of Si, Ge, GaP, GaAs, GaSb, InP, InAs, and InSb from 1.5 to 6.0 eV},
  author = {Aspnes, D. E. and Studna, A. A.},
  journal = {Phys. Rev. B},
  volume = {27},
  issue = {2},
  pages = {985--1009},
  numpages = {0},
  year = {1983},
  month = {Jan},
  publisher = {American Physical Society},
  doi = {10.1103/PhysRevB.27.985},
  url = {https://link.aps.org/doi/10.1103/PhysRevB.27.985}
}

@article{nSi_schinke,
    author = {Schinke, Carsten and Christian Peest, P. and Schmidt, Jan and Brendel, Rolf and Bothe, Karsten and Vogt, Malte R. and Kröger, Ingo and Winter, Stefan and Schirmacher, Alfred and Lim, Siew and Nguyen, Hieu T. and MacDonald, Daniel},
    title = "{Uncertainty analysis for the coefficient of band-to-band absorption of crystalline silicon}",
    journal = {AIP Advances},
    volume = {5},
    number = {6},
    pages = {067168},
    year = {2015},
    month = {06},
    issn = {2158-3226},
    doi = {10.1063/1.4923379},
    url = {https://doi.org/10.1063/1.4923379}
}

@article{Malitson,
author = {I. H. Malitson},
journal = {J. Opt. Soc. Am.},
number = {10},
pages = {1205--1209},
publisher = {Optica Publishing Group},
title = {Interspecimen Comparison of the Refractive Index of Fused Silica$\ast$,†},
volume = {55},
month = {Oct},
year = {1965},
url = {https://opg.optica.org/abstract.cfm?URI=josa-55-10-1205},
doi = {10.1364/JOSA.55.001205}
}

@article{nSi_franta_full_2017,
title = {Temperature dependent dispersion model of float zone crystalline silicon},
journal = {Applied Surface Science},
volume = {421},
pages = {405-419},
year = {2017},
issn = {0169-4332},
doi = {https://doi.org/10.1016/j.apsusc.2017.02.021},
url = {https://www.sciencedirect.com/science/article/pii/S0169433217303720},
author = {Daniel Franta and Adam Dubroka and Chennan Wang and Angelo Giglia and Jirí Vohánka and Pavel Franta and Ivan Ohlídal}
}

@misc{lithography,
    author = {Rochester Institute of Technology},
    title = {Optical Properties of Thin Films for DUV and VUV Microlithography},
    url = {https://www.rit.edu/kgcoe/microsystems/lithography/thinfilms/thinfilms/thinfilms.html},
    urldate={2025-07-15},
    year = {2012}
}

@article{sio2_properties2,
author = {Muamer Zukic and Douglas G. Torr and James F. Spann and Marsha R. Torr},
journal = {Appl. Opt.},
number = {28},
pages = {4284--4292},
publisher = {Optica Publishing Group},
title = {Vacuum ultraviolet thin films. 1: Optical constants of BaF2, CaF2, LaF3, MgF2, Al2O3, HfO2, and SiO2 thin films},
volume = {29},
month = {Oct},
year = {1990},
url = {https://opg.optica.org/ao/abstract.cfm?URI=ao-29-28-4284},
doi = {10.1364/AO.29.004284}
}

@article{sio2_properties1,
author = {Rei Kitamura and Laurent Pilon and Miroslaw Jonasz},
journal = {Appl. Opt.},
number = {33},
pages = {8118--8133},
publisher = {Optica Publishing Group},
title = {Optical constants of silica glass from extreme ultraviolet to far infrared at near room temperature},
volume = {46},
month = {Nov},
year = {2007},
url = {https://opg.optica.org/ao/abstract.cfm?URI=ao-46-33-8118},
doi = {10.1364/AO.46.008118}
}

@article{SiO2_dep_2001_khodier,
  title = {The effect of the deposition method on the optical properties of $\sio$ thin films},
  author = {S.A. Khodier and H.M. Sidki},
  journal = {Materials Science: Materials in Electronics},
  volume = {12},
  pages = {107-109},
  numpages = {12},
  year = {2001},
  doi = {10.1023/A:1011254220935},
  url = {https://link.springer.com/article/10.1023/A:1011254220935}
}

@article{excitonicBands_lxe_laporte_1977,
  title = {Evolution of excitonic bands in fluid xenon},
  author = {Laporte, P. and Steinberger, I. T.},
  journal = {Phys. Rev. A},
  volume = {15},
  issue = {6},
  pages = {2538--2544},
  numpages = {0},
  year = {1977},
  month = {Jun},
  publisher = {American Physical Society},
  doi = {10.1103/PhysRevA.15.2538},
  url = {https://link.aps.org/doi/10.1103/PhysRevA.15.2538}
}

@article{lxe_n_sinnockSmith,
  title = {Refractive Indices of the Condensed Inert Gases},
  author = {Sinnock, A. C. and Smith, B. L.},
  journal = {Phys. Rev.},
  volume = {181},
  issue = {3},
  pages = {1297--1307},
  numpages = {0},
  year = {1969},
  month = {May},
  publisher = {American Physical Society},
  doi = {10.1103/PhysRev.181.1297},
  url = {https://link.aps.org/doi/10.1103/PhysRev.181.1297}
}

@article{lar_lxe_GRACE2017,
title = {Index of refraction, Rayleigh scattering length, and Sellmeier coefficients in solid and liquid argon and xenon},
journal = {Nuclear Instruments and Methods in Physics Research Section A: Accelerators, Spectrometers, Detectors and Associated Equipment},
volume = {867},
pages = {204-208},
year = {2017},
issn = {0168-9002},
doi = {https://doi.org/10.1016/j.nima.2017.06.031},
url = {https://www.sciencedirect.com/science/article/pii/S0168900217306848},
author = {Emily Grace and Alistair Butcher and Jocelyn Monroe and James A. Nikkel}
}

@article{Heindl_LAr_2010,
doi = {10.1209/0295-5075/91/62002},
url = {https://dx.doi.org/10.1209/0295-5075/91/62002},
year = {2010},
month = {oct},
publisher = {},
volume = {91},
number = {6},
pages = {62002},
author = {Heindl, T. and Dandl, T. and Hofmann, M. and Krücken, R. and Oberauer, L. and Potzel, W. and Wieser, J. and Ulrich, A.},
title = {The scintillation of liquid argon},
journal = {Europhysics Letters}
}

@article{fuji_lxe_2015,
title = {High-accuracy measurement of the emission spectrum of liquid xenon in the vacuum ultraviolet region},
journal = {Nuclear Instruments and Methods in Physics Research Section A: Accelerators, Spectrometers, Detectors and Associated Equipment},
volume = {795},
pages = {293-297},
year = {2015},
issn = {0168-9002},
doi = {https://doi.org/10.1016/j.nima.2015.05.065},
url = {https://www.sciencedirect.com/science/article/pii/S016890021500724X},
author = {Keiko Fujii and Yuya Endo and Yui Torigoe and Shogo Nakamura and Tomiyoshi Haruyama and Katsuyu Kasami and Satoshi Mihara and Kiwamu Saito and Shinichi Sasaki and Hiroko Tawara}
}

@unpublished{index_cf4_french,
  TITLE = {{Les compteurs Cherenkov : applications et limites pour l'identification des particules. D{\'e}veloppements et perspectives}},
  AUTHOR = {Seguinot, J.},
  URL = {https://cel.hal.science/cel-00645583},
  NOTE = {Lecture},
  TYPE = {{\'E}cole th{\'e}matique},
  ADDRESS = {Maubuisson, (France), du 26-30 septembre 1988 : 7{\`e}me session},
  YEAR = {1988},
  MONTH = Sep,
  HAL_ID = {cel-00645583},
  HAL_VERSION = {v1},
}

@article{index_cf4_chemPhys,
    author = {Abbiss, C. P. and Knobler, C. M. and Teague, R. K. and Pings, C. J.},
    title = {Refractive Index and Lorentz—Lorenz Function for Saturated Argon, Methane, and Carbon Tetrafluoride},
    journal = {The Journal of Chemical Physics},
    volume = {42},
    number = {12},
    pages = {4145-4148},
    year = {1965},
    month = {06},
    issn = {0021-9606},
    doi = {10.1063/1.1695909},
    url = {https://doi.org/10.1063/1.1695909},
    eprint = {https://pubs.aip.org/aip/jcp/article-pdf/42/12/4145/18838441/4145_1_online.pdf},
}

@article{solar_multilayer_review,
title = {A review of anti-reflection and self-cleaning coatings on photovoltaic panels},
journal = {Solar Energy},
volume = {199},
pages = {63-73},
year = {2020},
issn = {0038-092X},
doi = {https://doi.org/10.1016/j.solener.2020.01.084},
url = {https://www.sciencedirect.com/science/article/pii/S0038092X20300918},
author = {Ali Samet Sarkın and Nazmi Ekren and Şafak Sağlam}
}

@article{uvMulti_Hamden2011,
  title = {Ultraviolet antireflection coatings for use in silicon detector design},
  volume = {50},
  ISSN = {1539-4522},
  url = {http://dx.doi.org/10.1364/AO.50.004180},
  DOI = {10.1364/ao.50.004180},
  number = {21},
  journal = {Applied Optics},
  publisher = {Optica Publishing Group},
  author = {Hamden,  Erika T. and Greer,  Frank and Hoenk,  Michael E. and Blacksberg,  Jordana and Dickie,  Matthew R. and Nikzad,  Shouleh and Martin,  D. Christopher and Schiminovich,  David},
  year = {2011},
  month = jul,
  pages = {4180}
}

@article{JIA_solarcells_broadband_2017,
title = {Preparation and properties of five-layer graded-refractive-index antireflection coating nanostructured by solid and hollow silica particles},
journal = {Thin Solid Films},
volume = {642},
pages = {174-181},
year = {2017},
issn = {0040-6090},
doi = {https://doi.org/10.1016/j.tsf.2017.09.038},
url = {https://www.sciencedirect.com/science/article/pii/S0040609017307113},
author = {Guiyu Jia and Zihan Ji and Hongning Wang and Ruoyu Chen}
}

@INPROCEEDINGS{multilayer_2005,
  author={Wright, D.N. and Marstein, E.S. and Holt, A.},
  booktitle={Conference Record of the Thirty-first IEEE Photovoltaic Specialists Conference, 2005.}, 
  title={Double layer anti-reflective coatings for silicon solar cells}, 
  year={2005},
  volume={},
  number={},
  pages={1237-1240},
  doi={10.1109/PVSC.2005.1488363}}

@misc{Alex_2024LoLX2,
  author       = {Xiang Li},
  title        = {The Light-only Liquid Xenon (LoLX) Experiment Phase 2 Study},
  year         = {2024},
  howpublished = {Oral presentation at the 2024 CAP Congress, Western University, Canada},
  month        = may,
  note         = {Presented on May 27, 2024},
}

@misc{sbc_snowmass2021,
      title={Snowmass 2021 Scintillating Bubble Chambers: Liquid-noble Bubble Chambers for Dark Matter and CE$\nu$NS Detection}, 
      author={E. Alfonso-Pita and M. Baker and E. Behnke and A. Brandon and M. Bressler and B. Broerman and K. Clark and R. Coppejans and J. Corbett and C. Cripe and M. Crisler and C. E. Dahl and K. Dering and A. de St. Croix and D. Durnford and K. Foy and P. Giampa and J. Gresl and J. Hall and O. Harris and H. Hawley-Herrera and C. M. Jackson and M. Khatri and Y. Ko and N. Lamb and M. Laurin and I. Levine and W. H. Lippincott and X. Liu and R. Neilson and S. Pal and J. Phelan and M. -C. Piro and S. Priya and S. Ray and E. Rich and Z. Sheng and A. Sloss and X. Struyk and E. Vázquez-Jáuregui and D. Velasco and S. Westerdale and T. J. Whitis and W. Zha and R. Zhang},
      year={2022},
      eprint={2207.12400},
      archivePrefix={arXiv},
      primaryClass={physics.ins-det},
      url={https://arxiv.org/abs/2207.12400}, 
}

@article{legend_result,
  title = {First results on the search for lepton number violating neutrinoless double-$\ensuremath{\beta}$ decay with the LEGEND-200 experiment},
  author = {al., H. Acharya et},
  journal = {Phys. Rev. Lett.},
  pages = {--},
  year = {2025},
  month = {Sep},
  publisher = {American Physical Society},
  doi = {10.1103/25tk-nctn},
  url = {https://link.aps.org/doi/10.1103/25tk-nctn}
}

@article{Herrera_vuv4_2024,
doi = {10.1088/1748-0221/19/08/T08003},
url = {https://dx.doi.org/10.1088/1748-0221/19/08/T08003},
year = {2024},
month = {aug},
publisher = {IOP Publishing},
volume = {19},
number = {08},
pages = {T08003},
author = {Hawley-Herrera, H. and Alfonso-Pita, E. and Behnke, E. and Bressler, M. and Broerman, B. and Clark, K. and Corbett, J. and Dahl, C.E. and Dering, K. and Croix, A.de St. and Durnford, D. and Giampa, P. and Hall, J. and Harris, O. and Lamb, N. and Laurin, M. and Levine, I. and Lippincott, W.H. and Liu, X. and Moss, N. and Neilson, R. and Piro, M.-C. and Pyda, D. and Sheng, Z. and Sweeney, G. and Vázquez-Jáuregui, E. and Westerdale, S. and Whitis, T.J. and Wright, A. and Wyman, E. and Zhang, R.},
title = {Batch VUV4 characterization for the SBC-LAr10 scintillating bubble chamber},
journal = {Journal of Instrumentation}
}

@article{carnesecchi_darkSide_light_2020,
	title = {Light detection in {DarkSide}-20k},
	volume = {15},
	issn = {1748-0221},
	url = {https://dx.doi.org/10.1088/1748-0221/15/03/C03038},
	doi = {10.1088/1748-0221/15/03/C03038},
	language = {en},
	number = {03},
	urldate = {2024-02-21},
	journal = {Journal of Instrumentation},
	author = {Carnesecchi, F.},
	month = mar,
	year = {2020},
	pages = {C03038}
}

@article{adhikari_nexo_2021,
	title = {{nEXO}: neutrinoless double beta decay search beyond 1028 year half-life sensitivity},
	volume = {49},
	issn = {0954-3899},
	shorttitle = {{nEXO}},
	url = {https://dx.doi.org/10.1088/1361-6471/ac3631},
	doi = {10.1088/1361-6471/ac3631},
	language = {en},
	number = {1},
	urldate = {2024-01-19},
	journal = {Journal of Physics G: Nuclear and Particle Physics},
	author = {Adhikari, G. and Kharusi, S. Al and et al.},
	month = dec,
	year = {2021},
	note = {Publisher: IOP Publishing},
	pages = {015104},
}

@article{PEN_Kuniak2019,
  title = {Polyethylene naphthalate film as a wavelength shifter in liquid argon detectors},
  volume = {79},
  ISSN = {1434-6052},
  url = {http://dx.doi.org/10.1140/epjc/s10052-019-6810-8},
  DOI = {10.1140/epjc/s10052-019-6810-8},
  number = {4},
  journal = {The European Physical Journal C},
  publisher = {Springer Science and Business Media LLC},
  author = {Kuźniak,  M. and Broerman,  B. and Pollmann,  T. and Araujo,  G. R.},
  year = {2019},
  month = mar 
}

@article{TPB_Francini_2013,
doi = {10.1088/1748-0221/8/09/C09010},
url = {https://dx.doi.org/10.1088/1748-0221/8/09/C09010},
year = {2013},
month = {sep},
publisher = {},
volume = {8},
number = {09},
pages = {C09010},
author = {R Francini and R M Montereali and E Nichelatti and M A Vincenti and N Canci and E Segreto and F Cavanna and F Di Pompeo and F Carbonara and G Fiorillo and F Perfetto},
title = {Tetraphenyl-butadiene films: VUV-Vis optical characterization from room to liquid argon temperature},
journal = {Journal of Instrumentation}
}

@Article{Pape_2024,
AUTHOR = {Pape, Sebastian and Fernández García, Marcos and Moll, Michael and Wiehe, Moritz},
TITLE = {Study of Neutron-, Proton-, and Gamma-Irradiated Silicon Detectors Using the Two-Photon Absorption–Transient Current Technique},
JOURNAL = {Sensors},
VOLUME = {24},
YEAR = {2024},
NUMBER = {16},
ARTICLE-NUMBER = {5443},
URL = {https://www.mdpi.com/1424-8220/24/16/5443},
PubMedID = {39205137},
ISSN = {1424-8220},
DOI = {10.3390/s24165443}
}

@misc{baudis_darwinxlzd_2024,
	title = {{DARWIN}/{XLZD}: a future xenon observatory for dark matter and other rare interactions},
	shorttitle = {{DARWIN}/{XLZD}},
	url = {http://arxiv.org/abs/2404.19524},
	doi = {10.48550/arXiv.2404.19524},
	urldate = {2025-06-10},
	publisher = {arXiv},
	author = {Baudis, Laura},
	month = apr,
	year = {2024},
	note = {arXiv:2404.19524 [astro-ph]},
}

@article{DUNE_updated,
doi = {10.1088/1748-0221/20/06/C06034},
url = {https://doi.org/10.1088/1748-0221/20/06/C06034},
year = {2025},
month = {jun},
publisher = {IOP Publishing},
volume = {20},
number = {06},
pages = {C06034},
author = {Botogoske, G. and on behalf of the DUNE collaboration},
title = {DUNE Photon Detection System},
journal = {Journal of Instrumentation}
}

@article{NV_zerophonon,
  title = {Assignment of the NV${}^{0}$ 575-nm zero-phonon line in diamond to a ${}^{2}E$-${}^{2}{A}_{2}$ transition},
  author = {Manson, N. B. and Beha, K. and Batalov, A. and Rogers, L. J. and Doherty, M. W. and Bratschitsch, R. and Leitenstorfer, A.},
  journal = {Phys. Rev. B},
  volume = {87},
  issue = {15},
  pages = {155209},
  numpages = {5},
  year = {2013},
  month = {Apr},
  publisher = {American Physical Society},
  doi = {10.1103/PhysRevB.87.155209},
  url = {https://link.aps.org/doi/10.1103/PhysRevB.87.155209}
}

@article{siBandgap_dmPaper,
    author = {Stanford, C. and Wilson, M. J. and Cabrera, B. and Diamond, M. and Kurinsky, N. A. and Moffatt, R. A. and Ponce, F. and von Krosigk, B. and Young, B. A.},
    title = {Photoelectric absorption cross section of silicon near the bandgap from room temperature to sub-Kelvin temperature},
    journal = {AIP Advances},
    volume = {11},
    number = {2},
    pages = {025120},
    year = {2021},
    month = {02},
    issn = {2158-3226},
    doi = {10.1063/5.0038392},
    url = {https://doi.org/10.1063/5.0038392},
    eprint = {https://pubs.aip.org/aip/adv/article-pdf/doi/10.1063/5.0038392/12991701/025120\_1\_online.pdf},
}

@article{siBandgap_solarCells,
title = {Absorption coefficient of silicon for solar cell calculations},
journal = {Solid-State Electronics},
volume = {22},
number = {9},
pages = {793-795},
year = {1979},
issn = {0038-1101},
doi = {https://doi.org/10.1016/0038-1101(79)90128-X},
url = {https://www.sciencedirect.com/science/article/pii/003811017990128X},
author = {K. Rajkanan and R. Singh and J. Shewchun}
}

@article{siAbsorption_Bucher,
    author = {Bücher, K. and Bruns, J. and Wagemann, H. G.},
    title = {Absorption coefficient of silicon: An assessment of measurements and the simulation of temperature variation},
    journal = {Journal of Applied Physics},
    volume = {75},
    number = {2},
    pages = {1127-1132},
    year = {1994},
    month = {01},
    issn = {0021-8979},
    doi = {10.1063/1.356496},
    url = {https://doi.org/10.1063/1.356496},
    eprint = {https://pubs.aip.org/aip/jap/article-pdf/75/2/1127/18663467/1127\_1\_online.pdf}
}

@article{siBandgap_cardona_1985,
  title = {Temperature dependence of band gaps in Si and Ge},
  author = {Lautenschlager, P. and Allen, P. B. and Cardona, M.},
  journal = {Phys. Rev. B},
  volume = {31},
  issue = {4},
  pages = {2163--2171},
  numpages = {0},
  year = {1985},
  month = {Feb},
  publisher = {American Physical Society},
  doi = {10.1103/PhysRevB.31.2163},
  url = {https://link.aps.org/doi/10.1103/PhysRevB.31.2163}
}

@article{photoAbs_doping_Abroug2008,
  title = {Optical and thermal properties of doped semiconductor},
  volume = {153},
  ISSN = {1951-6401},
  url = {http://dx.doi.org/10.1140/epjst/e2008-00386-7},
  DOI = {10.1140/epjst/e2008-00386-7},
  number = {1},
  journal = {The European Physical Journal Special Topics},
  publisher = {Springer Science and Business Media LLC},
  author = {Abroug,  S. and Saadallah,  F. and Yacoubi,  N.},
  year = {2008},
  month = jan,
  pages = {29–32}
}

@article{mcintyre_avalanche_1973,
	title = {On the avalanche initiation probability of avalanche diodes above the breakdown voltage},
	volume = {20},
	issn = {1557-9646},
	url = {https://ieeexplore.ieee.org/abstract/document/1477372},
	doi = {10.1109/T-ED.1973.17715},
	number = {7},
	urldate = {2023-10-12},
	journal = {IEEE Transactions on Electron Devices},
	author = {McIntyre, R.J.},
	month = jul,
	year = {1973},
	note = {Conference Name: IEEE Transactions on Electron Devices},
	pages = {637--641}
}

@article{mcintyre_microplasma,
    author = {McIntyre, R. J.},
    title = {Theory of Microplasma Instability in Silicon},
    journal = {Journal of Applied Physics},
    volume = {32},
    number = {6},
    pages = {983-995},
    year = {1961},
    month = {06},
    issn = {0021-8979},
    doi = {10.1063/1.1736199},
    url = {https://doi.org/10.1063/1.1736199},
    eprint = {https://pubs.aip.org/aip/jap/article-pdf/32/6/983/18323870/983_1_online.pdf},
}

@article{mcintyre_new_1999,
	title = {A new look at impact ionization-{Part} {I}: {A} theory of gain, noise, breakdown probability, and frequency response},
	volume = {46},
	issn = {1557-9646},
	shorttitle = {A new look at impact ionization-{Part} {I}},
	doi = {10.1109/16.777150},
	number = {8},
	journal = {IEEE Transactions on Electron Devices},
	author = {McIntyre, R.J.},
	month = aug,
	year = {1999},
	note = {Conference Name: IEEE Transactions on Electron Devices},
	pages = {1623--1631}
}

@article{chaves_2021_analyticAvalanche,
doi = {10.1088/1361-6641/ac00d0},
url = {https://doi.org/10.1088/1361-6641/ac00d0},
year = {2021},
month = {jul},
publisher = {IOP Publishing},
volume = {36},
number = {8},
pages = {085008},
author = {Chaves De Albuquerque, Tulio and Issartel, Dylan and Gao, Shaochen and Benhammou, Younes and Golanski, Dominique and Clerc, Raphaël and Calmon, Francis},
title = {An analytical solution for McIntyre’s model of avalanche triggering probability for SPAD compact modeling and performance exploration},
journal = {Semiconductor Science and Technology}
}

@ARTICLE{oldham_avalanche,
  author={Oldham, W.G. and Samuelson, R.R. and Antognetti, P.},
  journal={IEEE Transactions on Electron Devices}, 
  title={Triggering phenomena in avalanche diodes}, 
  year={1972},
  volume={19},
  number={9},
  pages={1056-1060},
  doi={10.1109/T-ED.1972.17544}
}

@article{Lewis_2025_qy,
   title={Measurements of the quantum yield of silicon using Geiger-mode avalanching photodetectors},
   volume={85},
   ISSN={1434-6052},
   url={http://dx.doi.org/10.1140/epjc/s10052-025-13883-x},
   DOI={10.1140/epjc/s10052-025-13883-x},
   number={2},
   journal={The European Physical Journal C},
   publisher={Springer Science and Business Media LLC},
   author={Lewis, Harry and Mahtab, Mahsa and Retière, Fabrice and De St. Croix, Austin and Raymond, Kurtis and Henriksson-Ward, Maia and Morrison, Nicholas and Zhang, Aileen and Capra, Andrea and Underwood, Ryan},
   year={2025},
   month=feb
}

\end{document}